\documentclass[12pt,a4]{article}
\usepackage{times,fancyhdr,bm}
\usepackage{url,color}

\usepackage{pdflscape}
\usepackage{epsfig}
\usepackage{subfig}    
\usepackage{natbib} 
\usepackage{setspace} 
\usepackage{enumerate} 
\usepackage{rotating} 
\usepackage{mathrsfs} 
\usepackage{amssymb} 
\usepackage{amsthm,amsfonts,amsbsy}
\usepackage{ctable}
\usepackage{scalerel}
\usepackage[intlimits]{amsmath} 
\usepackage{graphicx}
\usepackage{geometry}
\usepackage{multicol}
\usepackage{color}
\usepackage[normalem]{ulem}
\definecolor{Magenta}{rgb}{1,0,1}
\definecolor{Green2}{rgb}{0,0.5,0}
\definecolor{Ciano}{rgb}{0,0.75,0.75}
\definecolor{darkgreen}{rgb}{0.01, 0.6, 0.01}
\newcommand{\RA}{\textcolor{blue}} 
\newcommand{\RB}{\textcolor{black}} 

\newcommand*{\sdash}[1]{{\color{#1} \protect\rule[.5ex]{0.5ex}{1pt}}}
\newcommand*{\ldash}[1]{{\color{#1} \protect\rule[.5ex]{3.5ex}{1pt}}}
\newcommand*{\mdash}[1]{{\color{#1} \protect\rule[.5ex]{1.5ex}{1pt}}}
\newcommand{\ccirc}{\kern0.5ex\vcenter{\hbox{$\scriptstyle\bigcirc$}}\kern0.5ex}

\begin{document}
\graphicspath{{graphics/}}

\begin{center}
{\LARGE \textit{A posteriori} assessment of consumption speed correction for LES with tabulated methods}
\end{center}

\vspace{10pt}

\noindent I.~Langella$^{1,2}$, N. A. K. Doan$^{1,*}$\\
\\
$^{1}$Faculty of Aerospace Engineering, Delft University of Technology, Delft, Netherlands;
$\; ^2$Department of Aeronautical and Automotive Engineering, Loughborough University, Loughborough, UK\\

\vspace{1.5in}
\noindent$\rm{\underline{^* Corresponding~~author}}$: \\
Faculty of Aerospace Engineering, Delft University of Technology\\ 
Kluyverweg 1, 2629 HS Delft, The Netherlands\\
E-mail: n.a.k.doan@tudelft.nl \\ 

\vspace{20pt}
\begin{center}
(Draft to CST \today)\\
{\underline {Running Title:}} Consumption speed correction for LES
\end{center}

\newpage
\begin{abstract}
Large eddy simulation of propane/air jet flame in the wrinkled flamelets regime of the Borghi diagram is used to assess the performance of a recently  developed consumption speed correction model in premixed combustion.

The combustion is modelled using flamelet tabulation with a presumed probability density function. The investigated flame does not lie in the shear layer and is subjected to self-driven oscillations, which is ideal to test the model performance.
The flame behaviour is first discussed using an accurate simulation performed on a refined mesh requiring no correction.
The same modelling framework used on a coarser mesh is observed to overestimate the consumption speed, leading to a shorter flame. The application of the consumption speed correction leads to the recovery of the flame length and width observed for the refined mesh, demonstrating its effectiveness. An extended model for partially-premixed combustion is also proposed and preliminarily tested on a high-pressure combustor.


\vspace{1.5cm}
\noindent
\parbox{1.0\textwidth}{ {\em Keywords:} Large eddy simulation, premixed combustion, partially premixed combustion, presumed PDF, consumption speed}
\newpage

   
\end{abstract}

 
\section{Introduction}\label{sec:Intro}
Pressure to develop new-generation, environmentally friendly combustion devices 
in response to the global emission targets has pushed researchers and industry in the last decade towards the investigation of lean premixed combustion, as high efficiency and low emissions can be achieved in this condition~\citep{Correa1998}.
However, lean premixed flames are  sensitive to fluctuations of pressure and heat release, which make them prone to unwanted oscillations and instability in practical devices~\citep{LieuwenY2005}. These effects have to be understood to develop stable-burning, reliable devices operating in this combustion mode. While experimental investigations are crucial to achieve this understanding, they are limited by cost and difficulties in performing experimental measurements. 
As a consequence, industrial design cycles have been complemented with computational fluid dynamics (CFD) investigations.

Among the numerical approaches for turbulent reacting flows, large eddy simulation (LES) has the potential to accurately predict time-dependent phenomena at a relatively small cost. In a LES, turbulent scales are resolved down to a cut-off scale, $\Delta$, with models to represent the smaller, residual scales~\citep{Pope_book}. Since the flame thickness is very thin in premixed flames~\citep{PoinsotV_book} and usually smaller than $\Delta$ in practical LES, the combustion and its interaction with turbulence need to be modelled. Many different approaches have been proposed in the past for this interaction, 
which can be broadly categorised into flamelet and non-flamelet, or geometrical and statistical approaches~\citep{GicquelSP2012}.

The focus in this work is on presumed-PDF approaches with an unstrained flamelet database which have demonstrated a good accuracy in premixed and partially premixed combustion at a relatively cheap computational cost, see for example~\RB{\cite{GicquelSP2012,LangellaS2016,LangellaSWF2016,LangellaCSS2018,GaleazzoSWHCF2019,LangellaHBVSZ2020,ChenLBS2019}}.
In such approaches, the filtered reaction rate $\overline{\dot{\omega}}$ needed by the reaction equation is generally modelled as
\begin{equation}\label{eq:PDFclosure}
    \overline{\dot{\omega}} = \overline{\rho}\int_0^1 \frac{\dot{\omega}(\eta)}{\rho(\eta)} P(\eta)\; d\eta
\end{equation}
where $\dot{\omega}$ is the laminar reaction rate in ${\rm kg\,m^{-3}\,s^{-1}}$, while $\rho$ and $\overline{\rho}$ are the laminar and filtered density respectively. The variable $\eta$ represents the sample space for a reaction progress variable $c$, defined to vary monotonically from 0 in the reactants to 1 in the products, and $P(\eta)$ is the \RB{Favre}
subgrid scale (SGS) presumed PDF, and is \RB{often} parametrised with first and second order \RB{(Favre-filtered)} moments of the progress variable (see more details in Section~\ref{sec:Modelling_combustion}). 
\RB{Note that in LES context, $P(c)$ is not a probability density function in a strict statistical sense, as it represents an ensemble of realisations in space (at the SGS level) at one time~\citep{ChenDRLS2018}. This quantity, which is sometimes referred to as filtered density function or FDF, will be referred here as SGS PDF for convenience.
It is also worth noting that the reaction rate in Eq.~\eqref{eq:PDFclosure} can be integrated by parts to replace $P$ with its cumulative distribution function. This is particularly useful when approaching the BML limit~\citep{BrayLM1985} (bimodal PDF), since a presumed PDF which depends on second order moment would become undetermined~\cite{LipatnikovSHSI2021}, and integration errors may become strong as it approaches the limit. Note also, however, that in the LES of turbulent combustion this exact limit is hardly achievable as it would imply that the second moment of the parametrisation variable is entirely at the unresolved level.}

The subgrid PDF statistically represents SGS processes such as flame oscillations and wrinkling, which, in a practical LES where the filter size $\Delta$ is larger than the laminar flame thickness $\delta_{th}$, are not resolved. An often used shape for this SGS PDF is the $\beta$-PDF~\citep{DomingoVPH2005,MoureauDV2011,LecocqRCV2010,NambullyDMV2014}, which has shown good accuracy at various regimes of the Borghi diagram~\citep{Borghi1990} \RB{at least for fully resolved meshes ($\Delta \leq \delta_{th}$)~\citep{LibbyW2000,BrayCLS2006,GalpinNVACD2008,VremanOG2009,KollaS2010a,DarbyshireS2012,GicquelSP2012,KlapdorMKJ2013,TrisjonoKKP2014,DoniniBOG2015,VanOijenDBTBdG2016,OttinoFFBG2016,DoniniBOG2017,GaleazzoSWHCF2019,LangellaSP2016,LangellaCSS2018,LangellaHBVSZ2020,ChenLBS2019,NilssonLDSYB2019,NillsonYDLSB2019}}. 
Alternative approaches are the \RB{sum of Dirac $\delta$-functions~\citep{RibertCP2004,RobinMCDRB2008,DarbyshireSH2010} and the}
laminar flamelet PDF~\citep{Bray2016,SalehiBSG2013}. The latter was shown to perform better in the case of flame intermittency and coarser meshes with $\Delta>\delta_{th}$~\citep{JinGB2008,SalehiB2010,FiorinaVADGV2010,LecocqRCV2011}, \RB{and has the advantage of not relying on the progress variable second moment if opportune modifications are made~\citep{LipatnikovSHSI2021,LipatnikovS2021b,LipatnikovSHSI2020}}.
Nonetheless, at very turbulent regimes (high Karlovitz number and thickened reaction zone regimes), the $\beta$-PDF \RB{was shown to be advantageous} because of its ability to model a wider range of progress variable variance. 

\uline{The role of this variance in the subgrid PDF and reaction rate modelling needs further clarification}. 
The transport equation for the SGS variance $\sigma^2_c$ of a reacting progress variable $c$ can be expressed as
\begin{equation}\label{eq:cvar_1}
    \frac{D \sigma^2_c}{Dt} = \mathcal{D} + \mathcal{P} + \mathcal{R} - \mathcal{S}
\end{equation}
where $\mathcal{D}$, $\mathcal{P}$, $\mathcal{R}$ and  $\mathcal{S}$ represent respectively the effects, at the SGS level, of diffusion (molecular plus subgrid contribution), turbulent production, reaction and scalar dissipation rate (SDR). This equation is very important as it features the balance between reaction, turbulence and dissipation typical of premixed flames.
vanishes for $\Delta \rightarrow 0$, but is of leading order for $\Delta = \mathcal{O}(\delta_{th})$~\citep{LangellaS2016}, implying that for $\Delta\ge \delta_{th}$ the variance is not zero even at laminar or quasi-laminar conditions ($\mathcal{P}=0$). This is explained by the fact that the flame is filtered out on the LES mesh and can still have subgrid oscillations, whose probability to occur is taken into account by the presumed (SGS) PDF model via a non-zero (SGS) variance. \RB{It is worth to note that the accuracy of term $\mathcal{R}$ was assessed \textit{a priori} by~\cite{NilssonLDSYB2019}, who showed that the presumed PDF approach with $\beta$-function is able to reproduce this term with very good accuracy at various Karlovitz numbers spanning the Borghi diagram}.
%

\begin{figure}[tb]
   	\centering
	\includegraphics[width=0.5\columnwidth]{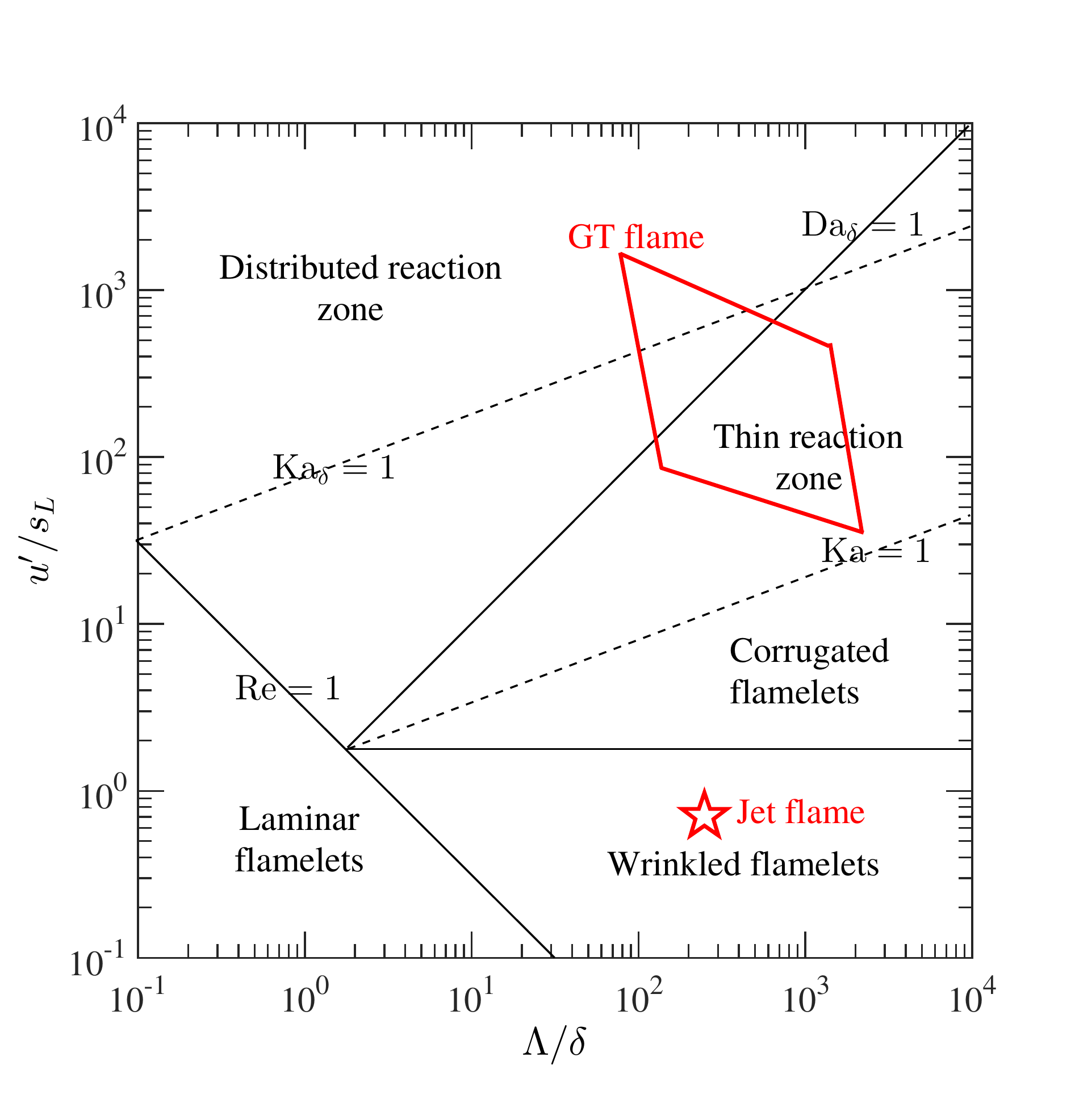}
	\caption{Premixed flame regime diagram with conditions for the premixed jet flame and the partially premixed GT flame investigated. The rms velocity, integral length scale, laminar flame speed and Zeldovich flame thickness are indicated with $u'$, $\Lambda$, $s_L$ and $\delta$ respectively.}	\label{fig:Borghi}
\end{figure}

\RB{The significance of the SGS variance with increasing LES filter width could also be interpeted in terms of its link with the filtered reaction rate. As $\Delta$ increases, in fact, increasingly more unburnt or burnt gases are taken into account in the filtering operation. Since $\dot{\omega}\approx 0$ outside the flame, the filtered reaction rate has to decrease, and this decrease is enforced by the increasing SGS variance in the presumed PDF approach, at least in qualitative terms. 
The non-zero SGS variance one would have in case of steady flame for filter widths larger than the laminar flame thickness, however, poses some challenge. Having in mind an unstretched 1D laminar flame, in absence of SGS wrinkling one can write~\citep{FiorinaVADGV2010}}
\begin{equation}\label{eq:Fiorina}
    \int_{-\infty}^{+\infty} \overline{\dot{\omega}}\; dx =  \int_{-\infty}^{+\infty} \dot{\omega}\; dx 
\end{equation}
\RB{This equation suggests that the overall burning rate before and after the filtering operation has to be preserved and, in more specific terms, being the described flame laminar, that the model used for the filtered reaction rate has to be able to reproduce the laminar flame speed. Nevertheless, as explained earlier the SGS variance, even if the exact one from DNS is used, is not zero for large $\Delta$ even in the case of zero-wrinkling; thus the equality in Eq.~\eqref{eq:Fiorina} would not hold in case of a presumed-PDF approach with dependency on second-order moments. Since $\mathcal{R}\rightarrow 0$ for $\Delta \rightarrow 0$ also implies $\sigma^2_c \rightarrow 0$ in Eq.~\eqref{eq:cvar} if no wrinkling is present ($\mathcal{P}=0$), for small filter widths Eq.~\eqref{eq:Fiorina} is implicitly verified as} the $\beta$-PDF degenerates into a $\delta$-function in this case. 
\RB{It was further shown  by~\cite{NillsonYDLSB2019} that the error on the burning speed is negligible up to $\Delta=\delta_{th}$, where $\delta_{th}$ is the laminar flame thickness. 
On the other hand, imaging now the case of a wrinkled flame, for increasing $\Delta$ the SGS variance has to increase to incorporate the physical effect of the wrinkling at the subgrid scale, so Eq.~\eqref{eq:Fiorina} would not hold in this case. Since the actual amount of SGS wrinkling is not known \textit{a priori} in a LES, imposing Eq.~\eqref{eq:Fiorina} may result in an incorrect burning speed and cannot thus be used to correct the burning speed at any condition. To overcome this limitation,}
the relation between modelled consumption speed and exact consumption speed was further analysed using DNS  by \cite{NillsonYDLSB2019}, where a more general equality valid for all variance values was \RB{proposed and} assessed: 
\begin{equation}\label{eq:HRRidentity}
    \int_{-\infty}^{+\infty} \overline{\dot{\omega}}\; dx =  \int_{-\infty}^{+\infty} \overline{\dot{\omega}}_e\; dx 
\end{equation}
In the above, the laminar reaction rate within the integral on the right-hand side of Eq.~\eqref{eq:Fiorina} is replaced by the `exact' filtered reaction rate, obtained as
\begin{equation}\label{eq:omega_ex}
    \overline{\dot{\omega}}_e = \int_{-\infty}^{+\infty} \dot{\omega}\, G(x; \Delta)\; dx
\end{equation}
where the implicit filter shape $G$ is approximated with a Gaussian filter. \RB{Equation~\eqref{eq:omega_ex} can thus be computed \textit{a priori} and then substituted into Eq.~\eqref{eq:HRRidentity} to obtain the right consumption speed independently of the variance value, and used to enforce the correct consumption speed in the presumed (SGS) PDF approach (left-hand side of Eq.~\eqref{eq:HRRidentity}). In practical term,}
to ensure the above equality using the $\beta$-PDF approach, a correction factor is required. This is necessary in cases with large $\Delta$, because the LHS of Eq. \eqref{eq:HRRidentity} \textcolor{Green2} is observed to overestimate the consumption speed~\citep{NillsonYDLSB2019}. In terms of mean reaction rate, similar overestimations as compared to DNS data were observed in previous studies~\citep{BrayCLS2006} where the $\beta$-PDF was used.
This behaviour is explained here as the result of two competing factors: on the one hand, a larger $\Delta$ implies larger SGS variance (due to larger probability of SGS fluctuations in the presumed-PDF approach), which decreases the overestimation; 
\uline{on the other hand, the integration of this modelled reaction rate over a larger cell (as a result of a coarse mesh) ultimately yields an overall overestimation of the consumption speed. The latter is particularly relevant for finite volume approaches where the reaction rate is treated as a source term in tabulated approaches and directly multiplied by the local cell volume}, \RB{and thus should be taken into account even when the SGS PDF is `exact'}.
The correction factor proposed by \cite{NillsonYDLSB2019} to remedy this was successfully assessed \textit{a priori} for a range of Karlovitz numbers, but was never demonstrated in an \textit{a posteriori} context.
The quasi-laminar jet flame configuration studied experimentally in~\cite{FurukawaYW2016} 
is thus chosen for this assessment. This flame lies in the wrinkled reaction zone regime of the Borghi diagram, see Fig.~\ref{fig:Borghi}, where the $\beta$-PDF is expected to lead to over-prediction of the flame speed. Thus, this configuration is particularly well suited for this investigation. A second configuration indicated in Fig.~\ref{fig:Borghi} is also tested to assess the model correction at a highly turbulent and partially premixed configuration.

%
%

The objective of this paper is to assess the effectiveness of the consumption speed correction proposed by \cite{NillsonYDLSB2019} by analysing the improvements in accuracy when using this correction from a base inaccurate simulation. To achieve this, the analysis is split into three steps.
First, a base case simulation of the quasi-laminar jet flame configuration of \cite{FurukawaYW2016} is performed on a fine grid, where $\Delta \approx \delta_{th}$. Second, a coarser mesh is employed (the base inaccurate simulation to be improved) to show that the flame speed spuriously increases as a consequence of $\Delta>\delta_{th}$, resulting in a shorter flame.
Third, the flame speed correction model is employed on the coarser grid to assess its effectiveness.
A generalised model for partially premixed combustion is then derived, which is proved to be independent of fuel, initial conditions, and easy to implement in CFD codes. The effectiveness of this generalised correction is preliminarily tested for a gas turbine (GT) combustor configuration at high pressure, where $\Delta\gg \delta_{th}$. Although only qualitative conclusions could be derived for the latter case, this study is a first attempt to provide evidence of the applicability of the correction to practical GT cases.

This paper is organised as follows. In Section~\ref{sec:Modelling}, the combustion model, flame speed correction and LES details are presented. The quasi-laminar jet flame configuration, mesh details and boundary conditions are described in Section~\ref{sec:CaseStudy}. Results for the jet flame are presented in Section~\ref{sec:Results}. In this section, the general flame structure and the interplay with the surrounding vortex dynamics are first discussed to characterise the flame behaviour in the quasi-laminar regime, and then the aforementioned analysis on consumption speed is conducted. In Section~\ref{sec:PPflame_description} and \ref{sec:PPflame_testcase}, the generalised flame speed correction model and its application on a real GT case at high pressure are discussed. Summary and conclusions are given in Section~\ref{sec:Conclusions}.
\section{Modelling details} \label{sec:Modelling}

  \subsection{Premixed combustion modelling} \label{sec:Modelling_combustion}
 An unstrained flamelet combustion model is employed to close the subgrid turbulence-combustion interaction in the LES.
 The effect of sub-grid wrinkling is taken into account statistically by using a presumed PDF approach, as discussed in Section~\ref{sec:Intro}. 
 \RB{\textit{A priori} analyses of the various modelling elements within this approach were performed in previous studies~\citep{DoanSC2017,LangellaDSP2018,NilssonLDSYB2019,NillsonYDLSB2019}}.
 Although the effect of strain may be important in premixed flames, recent works have found that, in the context of LES, this effect is predicted at the resolved level when most of energy is resolved~\citep{DoanSC2017,LangellaS2016}, \uline{at least for conditions away from blow-off}~\citep{ChenLBS2019}. 
 It is worth noting that these studies concern atmospheric flames at conditions which may not be fully representative of high-pressure configurations as the one discussed in Section~\ref{sec:PPflame_testcase}, and so one has to be mindful as subgrid strain effect could be more relevant in this case. On the other hand, satisfying comparisons between numerical and experimental data in previous studies of practical combustors (e.g. see \cite{LangellaCSS2018,LangellaHBVSZ2020}) suggest that the above energy criterion is still reasonable.
 \RB{Moreover, the conditions investigated in this work do not involve the intensive stretch regimes leading to significant alteration of the flame internal structure by turbulence (e.g. as shown for fan-stirred reactors in~\citep{AbdelB1985,AbdelBL1987}), and it was further demonstrated in~\citep{ChenLBS2019,SoliLC2021} that the used presumed PDF approach for LES does perform well up at least to moderate strain leading to formation of localised flame extinctions.}
 These reasons allow us to keep this simpler modelling framework, thus no subgrid strain modelling is used here. 
 
 
 The flamelet assumption assumes the flame to be thin enough so that small turbulent eddies wrinkle but do not modify its inner structure. The thermochemistry can thus be computed \textit{a priori} and linked to the LES by tracking the reaction progress.
In this work, this is done by defining a scaled progress variable based on products as
 $c = \Psi/\Psi^b$, where $\Psi = Y_{\rm CO_2}+Y_{\rm CO}$, $Y_k$ being the mass fraction of species $k$, and $\Psi^b$ the adiabatic value of $\Psi$. The choice of this definition is based on previous works~\citep{FiorinaBGTCD2003,LangellaSP2016,LangellaCSS2018,ChenLBS2019}.
 The transport equation for the filtered progress variable $\widetilde{c}$ is  
\begin{equation} \label{eq:progvar}
    \overline{\rho}\frac{D \widetilde{c}}{D t} =   
    \nabla\cdot \Bigl[\Bigl(\overline{\rho{\cal D}}+
       \frac{\mu_t}{Sc_t}\Bigr) \nabla\widetilde{c}\Bigr]+\overline{\dot{\omega}}_c
\end{equation}
 where $D/Dt$ refers to the total derivative, 
 $\overline{\rho D} \approx \mu/{\rm Sc} $ is the  filtered molecular diffusion term which is expressed \RB{using a gradient hypothesis} in terms of dynamic viscosity $\mu$ and Schmidt number ${\rm Sc}=0.7$, 
 $\mu_T$ is the SGS viscosity, ${\rm Sc}_t \approx 0.7$ is the SGS Schmidt number. \RB{It is worth to note that, although counter-gradient effects may exist in the configuration described Sec.~\ref{sec:CaseStudy}~\citep{VeynanteTBM1997}, they are implicitly captured at resolved level in a LES and the influence of unresolved fluxes on the main solution is small \citep{FiorinaVC2015}. Moreover, counter-gradient effects were observed to be strong at low Lewis number, but become milder as the Lewis number increases~\cite{LibbyB1981,ChakrabortyC2009,KleinCP2016}. Even using larger mesh sizes where the SGS contribution increases, the numerical truncation error in the LES was observed to be comparable to the SGS gradient effect~\citep{AllauddinKPC2017,LysenkoE2018}, and no significant differences were observed between the gradient and counter-gradient~\cite{ClarkFR1979} subgrid closures.}
 
 The filtered reaction rate of $c$, $\overline{\dot{\omega}}_c$, is is closed using a presumed-PDF approach as indicated in Eq.~\eqref{eq:PDFclosure}, and using a $\beta$-function, $P(\eta) = \beta(\eta;\,\widetilde{c},\sigma^2_c)$. This requires a value for the SGS variance of $c$, which is found in the LES using its transport equation:
 \begin{align}
    \overline{\rho}\frac{D \sigma_c^2}{D t}  \approx \nabla\cdot \Bigl[\Bigl(\overline{\rho{\cal D}}+
    \frac{\mu_t}{Sc_t}\Bigr) \nabla\sigma_c^2\Bigr]
	 + 2\,\frac{\mu_t}{Sc_t}\left(\nabla\widetilde{c}\cdot\nabla\widetilde{c}\right)
	 + 2\,\left(\overline{c\,\dot{\omega}_c}-\widetilde{c}\,\overline{\dot{\omega}}_c\right)
	 -2\overline{\rho}\,\widetilde{\varepsilon}_c\label{eq:cvar}
\end{align}
The above equation is formally identical to Eq.~\eqref{eq:cvar_1}. 
The reaction term is modelled in a manner similar to Eq.~\eqref{eq:PDFclosure}, where the integrand is computed \textit{a priori} for a one-dimensional, freely propagating, unstrained laminar premixed flame. 

The scalar dissipation rate (SDR, last term in Eq.~\eqref{eq:cvar}) has to balance both turbulent production and reaction terms~\citep{LangellaS2016}. One common approach to close this term is to use a linear relaxation model~\citep{Pitsch2006}:
\begin{equation}\label{eq:linearRelax}
    \overline{\rho}\widetilde{\varepsilon}_c =  \frac{\mu_t}{C_\alpha \Delta^2}\sigma^2_c
\end{equation}
with $C_\alpha$ being the model constant. This model was originally derived to balance the turbulent production term in Eq.~\eqref{eq:cvar} and is thus expected to perform better for conditions where the latter is strong compared to the reactive term. 
Since the reactive term $\mathcal{R}$ is expected to be of leading order for the quasi-laminar jet flame used in this work, at least for $\Delta$ of order or bigger than $\delta_{th}$ 
one can alternatively assume that the SDR of $\widetilde{c}$ consists of a contribution balancing the turbulent production and one balancing the reactive source, $\widetilde{\varepsilon}\approx \widetilde{\varepsilon}_{\rm prod}+\widetilde{\varepsilon}_{\rm reac}$ and then to further assume that the second contribution is directly proportional to the reaction term, $\overline{\rho}\widetilde{\varepsilon}_{\rm reac} = a\left(\overline{c\,\dot{\omega}_c}-\widetilde{c}\,\overline{\dot{\omega}}_c\right)$, \RB{where $a$ is a proportionality factor.} 
In the limit of $a \rightarrow 1$, \RB{considered here for simplicity,}
these two terms perfectly balance each other, \uline {and the application of this model is equivalent to using a SGS variance equation where $\widetilde{\varepsilon}_{\rm reac} 
-\mathcal{R}=0$}. 
A further alternative is to use the model proposed by~\cite{DunstanMCS2013}, which accounts for an explicit dependence on the subgrid kinetic energy $k$ and the thermochemical parameters: 
\begin{equation} \label{eq:SDRmodel}
    \widetilde{\varepsilon}_c = \left[1-\exp\left(\frac{-0.75\Delta}{\delta_{th}}\right)\right] 
    \left[\left(2K_c -\tau C_4\right)\frac{s_L}{\delta_{th}}+ C^{\prime}_3\frac{\epsilon_k}{k}\right]
    \frac{\sigma^2_c}{\beta_c}
\end{equation}
%
This expression has been tested in many past LES studies (e.g. see~\cite{LangellaS2016,LangellaSWF2016,LangellaSP2016}) and the reader is referred to one of these for further details.
\RB{Note that $\widetilde{\varepsilon}_c$ represents a subgrid quantity and thus has to approach zero for $\Delta \rightarrow 0$.}
The three different choices for the SDR of $\widetilde{c}$ are summarised in Table~\ref{tab:SDR} for clarity.
%
\uline{It is to note that dynamic evaluations of the model constants} \RB{($\beta_c$ in case of model C as it is the only tuneable constant in Eq.~\eqref{eq:SDRmodel}~\citep{DunstanMCS2013})}
\uline{are not possible for the configuration of Section~\ref{sec:CaseStudy} due to the quasi-laminar nature of the reacting flow}, \RB{as a sufficient range of scales for scale-similarity to hold is not present}~\citep{LangellaSWF2016}. 
This introduces additional uncertainty. In fact, while the value of the model constant for models A and B according to past works is within a narrow range, the constant $\beta_c$ for model C was shown to span a relatively large range~\citep{DunstanMCS2013,LangellaSP2016,LangellaSGC2015,LangellaCSS2018,ChenLBS2019}, 
which is partially caused by the closure for $k$ needed in Eq.~\eqref{eq:SDRmodel} and is certainly a disadvantage compared to models A and B. Due to the above reasons, the following choice is made for the current paper: model C is chosen for the fully premixed case to be examined in Section~\ref{sec:Results} due to the underlying assumptions within this model to better represent this quasi-laminar configuration for $\Delta\geq \delta_{th}$ ($\mathcal{R}\gg \mathcal{P}$). In this case,
the model limitation in finding an optimal value for the model constant are compensated by the availability of experimental data, which allow for a precise estimation of this value. It is, however, important to recognise that the lack of generality for the value of the constant would make the choice of model C less obvious for similar configurations in absence of validation data. Indeed,
model B will be employed instead for the partially premixed GT configuration of Section~\ref{sec:Results_2} as its underlying assumption are better verified in this case ($\mathcal{R}\sim \mathcal{P}$), and to limit the uncertainty in the choice of the model constant as no quantitative validation data is available in this case. Note that dynamic evaluation of the model constant in the GT configuration is again not performed in order to ensure that differences observed in the results obtained with and without the application of the consumption speed correction are a direct effect of this correction and not influenced by the dynamic constant algorithm.

     \subsection{Partially premixed combustion modelling}
The combustion model for partially premixed flames required for the GT configuration of Section~\ref{sec:PPflame_testcase} is briefly described here. Instead of a single flamelet for a specific equivalence ratio, a set of one-dimensional flamelets is computed for a number of equivalence ratios within the flammability limits. Note that a discussion on the use of premixed, rather than diffusion, flamelets to map the ensemble of reacting states is not the focus of this paper and this discussion can be found elsewhere, see for example~\cite{LangellaHBVSZ2020,ChenLBS2019,PierceM2004}.
An unscaled form of the progress variable, $c=\Psi$ is used as opposed to the scaled formulation employed for the premixed combustion, as this definition avoids the appearance of further terms to be modelled in the progress variable equation~\citep{LangellaCSS2018,LangellaHBVSZ2020,ChenLBS2019}. This definition was shown to introduce some numerical errors that can affect predictions in cases with local extinctions~\citep{ChenLBS2019}. However, local extinctions are not expected in the configurations studied in this study, and this formulation is thus sufficient. 

The analytical form of the progress variable equation and its SGS variance is identical to that of Eqs.~\eqref{eq:progvar} and \eqref{eq:cvar} given the definition of the progress variable used here.
To track the different mixing states in partially-premixed flames, a mixture fraction variable $\xi$ is  needed in addition to the progress variable, whose Favre-filtered transport equation is:
\begin{equation} \label{eq:Mixfrac}
 \overline{\rho}\frac{D \widetilde{\xi}}{D t} = 
      \nabla\cdot \Bigl[\Bigl(\overline{\rho \mathcal{D}}+
       \frac{\mu_t}{Sc_t}\Bigr) \nabla\widetilde{\xi}\Bigr]\RA{.}
 \end{equation}
The model as it has been formulated is known as flamelet generated manifold or FGM~\citep{VanOijenDBTBdG2016}, and is used here in conjunction with a presumed-PDF approach to statistically account for the subgrid turbulence-flame interaction. 
The filtered reaction rate in Eq.~\eqref{eq:progvar} is closed as
\begin{equation} \label{eq:PPclosure}
    \overline{\dot{\omega}}_c \; = \; \int^1_0\int_0^1 \dot{\omega}(\zeta, \eta)\;
    P(\zeta,\eta) \; d\zeta\;d\eta,
\end{equation}
where $\zeta$ is the sample space for $\xi$ and $P(\zeta,\eta)$ is the SGS joint PDF. This PDF is expressed as 
$P(\zeta,\eta) = P(\zeta)\,P(\eta|\zeta)$ using Bayesian rule,
and the PDFs of $c$ and $\xi$ are 
modelled using Beta functions~\citep{Abramo70}. A value for the SGS variance of $\widetilde{\xi}$, $\sigma^2_\xi$, is required, which is computed by solving its transport equations:
\begin{align}
    \overline{\rho}\frac{D \sigma_\xi^2}{D t} \approx \nabla\cdot \Bigl[\Bigl(\overline{\rho{\cal D}}+
    \frac{\mu_t}{Sc_t}\Bigr) \nabla\sigma_\xi^2\Bigr] 
	 + 2\,\overline{\rho}\,\frac{\nu_t}{Sc_t}\left(\nabla\tilde{\xi}\cdot\nabla\tilde{\xi}\right)
	 - 2\,\overline{\rho}\,\widetilde{\varepsilon}_\xi \label{eq:zvar}
\end{align}
where the SDR of $\widetilde{\xi}$ \uline{is modelled with the linear relaxation model}, \RB{(same functional form of Eq.~\eqref{eq:linearRelax})},
which is appropriate as $\widetilde{\xi}$, unlike the progress variable, is a passive scalar.
\RB{The value of the model constant, $C_\alpha=2$~\citep{Pitsch2006}, is not tuned in this equation as it was shown to perform well for LES in moderate to high turbulent flow conditions~\citep{ChenLBS2019,SoliLC2021,LangellaCSS2018,LangellaHBVSZ2020}. However, the reader is reminded that a recent~\textit{a priori} analysis~\citep{SitteAGMC2021} has indicated that this value can be significantly higher at certain conditions}.

\subsection{LES details}\label{sec:Modelling_LES}
The Favre-filtered transport equations for mass, momentum and specific enthalpy $\widetilde{h}$ (defined here as the sum of sensible, $\widetilde{h}_s$ and formation, $\widetilde{\Delta h}_f^0$, enthalpies), along with the equations for combustion described in Section~\ref{sec:Modelling_combustion}, are solved using the low-Mach formulation of the reacting Navier-Stokes equations and the finite volume method. An additional transport equation for a passive scalar tracker, $\widetilde{Z}$, is used (for the premixed flame case only) to take into account the mixing with air in non-reacting regions, as will be explained later.
The pressure-density-velocity coupling is solved using the PISO loop~\citep{Bressloff2011} implemented in OpenFOAM~\citep{Weller1998}, and the equations loop is iterated 5 times per each time step. The subgrid viscosity in the momentum equation is modelled using a one-equation model for the subgrid kinetic energy $k$, which is treated as in previous 
works~\citep{ChaiM2010,LangellaCSS2018,ChenLBS2019} and where the pressure-work term is neglected.
The turbulent transport term in all scalar transport 
equations is  modelled using a gradient hypothesis. The molecular dynamic viscosity is temperature dependent via Sutherland's law as in~\cite{LangellaCSS2018}. 

The temperature is computed as $\widetilde{T} = T_0 + (\widetilde{h}- \widetilde{\Delta h}_f^0)/\widetilde{C}_p^*$,
where $T_0=298.5\,{\rm K}$ 
and $C_p^* = (\int_{T_0}^T C_p \, dT')/(T-T_0)$ is precomputed. 
This equation comes immediately by inverting $h = \Delta h^0_f + \int_{T_0}^T C_p \, d \, T$ and applying the theoreom of the integral mean. 
The mixture density is computed from the state equation as $\overline{\rho} =  \overline{p}\widetilde{W}/(R_0 \widetilde{T})$, where $\overline{p}$ is the modified pressure~\citep{Pope_book}, $\widetilde{W}$ is the mixture molecular weight
and $R_0$ is the universal gas constant.  The Favre-filtered values for
$\widetilde{\Delta h}_f^0$,
$\widetilde{C}_p^*$ and 
$\widetilde{W}$ are obtained from the flamelets database using an equation consistent with Eq.~\eqref{eq:PDFclosure} (Eq.~\eqref{eq:PPclosure} for partially premixed flames), and then tabulated in terms of the controlling variables
$\widetilde{c}$ and $\sigma^2_c$ (also $\widetilde{\xi}$ and $\sigma^2_\xi$ for partially premixed flames). 
In the premixed flame model, to account for the air mixing effect on the thermodynamic properties, the generic thermodynamic quantity $\Phi_{\rm reac}$, which can be the specific heat capacity for example, from the precomputed table is weighted with the value in the air, $\Phi_{\rm air}$, using a bimodal expression 
\begin{equation*}
    \widetilde{\Phi} = \widetilde{Z}\Phi_{\rm reac} + (1-\widetilde{Z})\Phi_{\rm air}
\end{equation*}
Second order central schemes are used for the convective term in all transported variables,
with TVD limiters used for the scalars to deal with steep gradients in the flame region. An implicit
Euler scheme is used for time marching as it was observed to be of similar accuracy as second order schemes in OpenFOAM~\citep{LangellaCSS2018} but more favorable in terms of numerical instability issues. 
The flamelet database is computed using the code
Chem1d~\citep{Chem1d_code} with detailed chemistry for propane~\citep{KeeGSM1985} (the kerosene mechanism of~\cite{DagautC2006} for the partially premixed flame of Section~\ref{sec:PPflame_testcase}). This also provides the computed values for laminar flame speed and laminar flame thickness of 0.34 m/s and 0.4 mm respectively. 

Note that the quasi-laminar behaviour of the flame to be discussed in Section~\ref{sec:CaseStudy} may imply the presence of local variation of equivalence ratio and/or thermo-diffusive instabilities due to differential diffusion~\citep{BisettiCCH2009,DineshSLOT2016}, which in turn can be coupled to heat losses effects~\citep{MercierAMDGVF2014,DoniniBOG2015}. 
The effect of preferential diffusion on the progress variable reaction rate and the thermochemical parameters needed for Eq.~\eqref{eq:SDRmodel} was tested for the propane/air flamelet at $\phi=0.85$ using detailed computations in Chem1d~\citep{Chem1d_code} and found to be negligible. Also, no observation regarding differential diffusion  was made in the experiments~\citep{FurukawaYW2016}. Since
the treatment of preferential diffusion is inessential for the analysis on burning speed correction to be conducted in Section~\ref{sec:Results} and it would even add uncertainty due to the required subgrid modelling for $\Delta>\delta_{th}$~\citep{NambullyDMV2014}, it is
not taken into account in the laminar flamelet model used here.
On the other hand, heat losses may have a more substantial influence on the flame behaviour near the anchoring point. In fact, although the generic effect on temperature is taken into account (by transporting the enthalpy equation and using non-adiabatic walls), possible effects on the reaction rate are not (similarly to what done in~\cite{LangellaSWF2016}). Despite this limitation, this modelling is avoided to limit the uncertainty in the evaluation of the flame speed correction under investigation here. Nonetheless, the generalities of the analysis to be conducted for the jet flame configuration are not violated as will be discussed in Section~\ref{sec:Results}.

\subsection{Flame speed correction}\label{sec:Modelling_FlameSpeed}
As discussed in the Introduction section, for $\Delta > \delta_{th}$ the presumed PDF is unable to mimic the correct SGS distribution and Eq.~\eqref{eq:HRRidentity} does not hold. Consequently, for values of $\sigma^2_c>0$ but laminar or quasi-laminar conditions, the correct filtered laminar flame speed is not guaranteed even when $\sigma^2_c$ is estimated exactly. For a one-dimensional flame, Eq.~\eqref{eq:HRRidentity} can be enforced by introducing the filter-dependent scaling factor $f$~\citep{NillsonYDLSB2019}:
\begin{equation}\label{eq:fdelta}
    f(\Delta) = \frac{\int_{-\infty}^{+\infty}\overline{\dot{\omega}}_e(Y_k,T;\,\Delta) \; dx}{\int_{-\infty}^{+\infty}\overline{\dot{\omega}}(\overline{c},\sigma^2_c;\,\Delta) \; dx}
\end{equation}
in which $\overline{\dot{\omega}}$ is the modelled rate computed from Eq.~\eqref{eq:PDFclosure} and the `exact' filtered reaction rate, $\overline{\dot{\omega}}_e$, is computed for the one-dimensional flame with detailed chemistry and then applying a Gaussian filter. For the one dimensional flame the corrected reaction rate is then obtained by multiplying the modelled rate by $f$ corresponding to the local value of $\Delta$. For a three-dimensional flame, the `exact' value $\overline{\dot{\omega}}_e$ is unknown. However, if the local flame structure is one-dimensional, which is an assumption already used for flamelet models, one can also pre-compute $f(\Delta)$ from the one-dimensional flame and store it in a look-up table. By combining Eqs.~\eqref{eq:HRRidentity} and \eqref{eq:fdelta}, and substituting the one-dimensional integral with a volume integral, the equation to impose in the LES becomes:
\begin{equation}\label{eq:CorrectedOmega}
    \int_\mathcal{V} \overline{\dot{\omega}}_e \; d\mathcal{V} = 
    \int_\mathcal{V} f\overline{\dot{\omega}}\; d\mathcal{V}
\end{equation}
where $\mathcal{V}$ is the local cell volume and $\Delta$ is approximated here as $\mathcal{V}^{1/3}$. 
As the volume integral of the reaction rate is directly proportional to the consumption speed, the described correction is for the consumption speed, and in principle not for the reaction rate, which is the reason why the same scaling is not applied to the reaction term $\mathcal{R}$ in Eq.~\eqref{eq:cvar_1} (or equivalently Eq.~\eqref{eq:cvar}). Thus the proposed correction does not affect the SGS variance, which in turn would affect the reaction rate in Eq.~\eqref{eq:CorrectedOmega}.
Moreover, limiting the reactive term in the SGS variance would imply altering the intricate balance between reaction, turbulent production and the scalar dissipation rate (respectively terms $\mathcal{R}$, $\mathcal{P}$ and $\mathcal{S}$ in Eq.~\eqref{eq:cvar_1}) in premixed combustion. Thus, one should account for this coupling first, including a revision of the modelling for the scalar dissipation rate, which is beyond the scope of the present work. A sensitivity study was conducted on the quasi-laminar configuration of Section~\ref{sec:CaseStudy}, where the correction of Eq.~\eqref{eq:CorrectedOmega} was applied also to the reactive term $\mathcal{R}$ in the SGS variance equation, showing indeed that this choice leads only to partial recovery of the flame length. For these reasons the investigation of the role of the SGS variance and the coupling between $\mathcal{R}$ and the subgrid SDR in it, is postponed to a future work, although some considerations are provided at the end of Section~\ref{sec:Results_3}.
For the same reasons additional corrections to account for a change in diffusion (e.g. see \cite{ColinDVP2000,Charlette2002}), are not applied here. It is worth noting that the proposed correction is applied at no extra cost for the LES.

\begin{figure}[tb]
   	\centering
	\includegraphics[width=0.5\columnwidth]{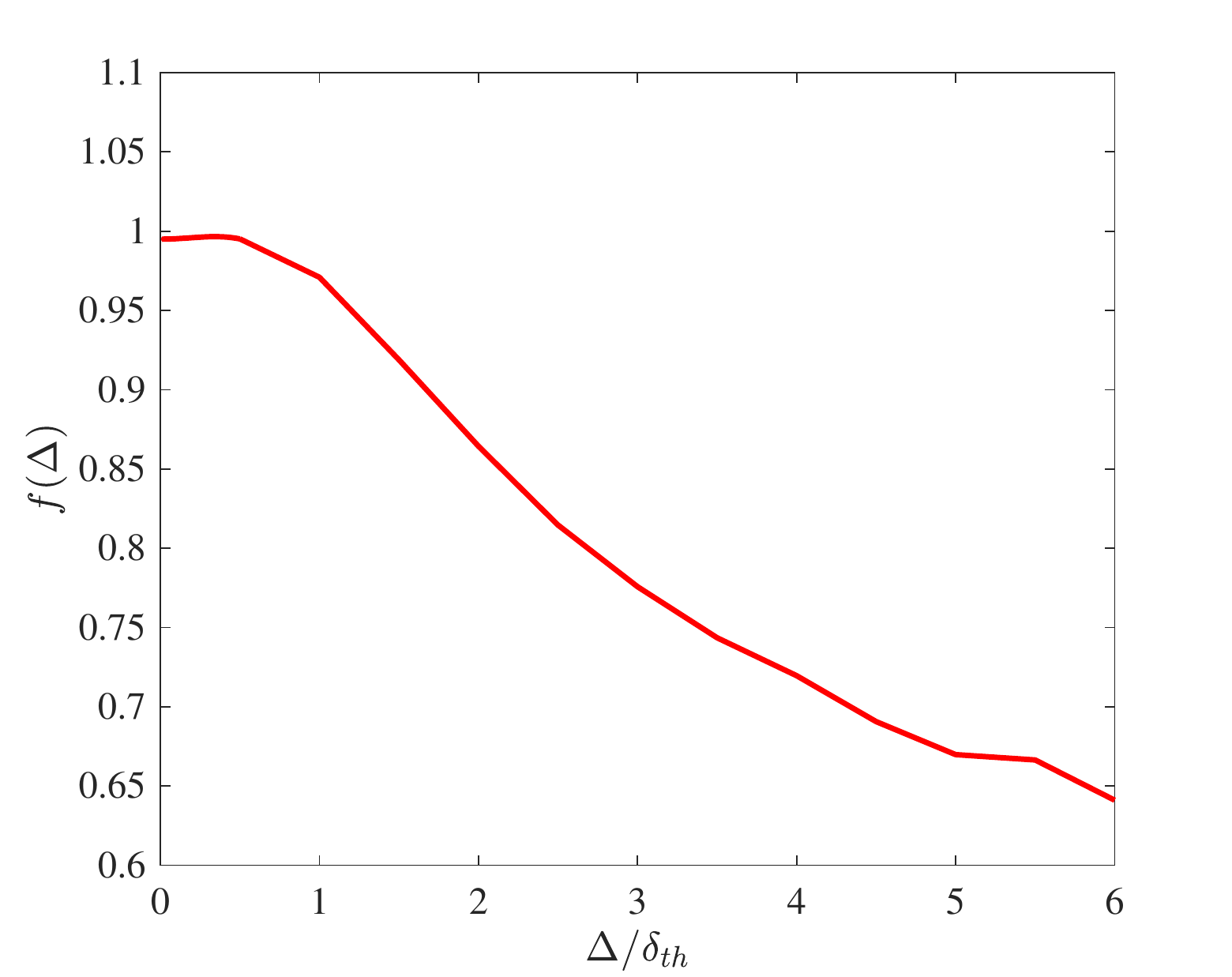}
	\caption{Scaling factor, Eq.~\eqref{eq:fdelta}, for various values of $\Delta/\delta_{th}$ for a propane/air flame at equivalence ratio $\phi=0.85$.}	\label{fig:fdelta_propane}
\end{figure}

The scaling factor $f(\Delta)$ is shown for propane/air mixture and equivalence ratio $\phi=0.85$ in Fig.~\ref{fig:fdelta_propane}, as this is the condition of the flame to be analysed in Section~\ref{sec:Results}.
%
For small values of $\Delta$ and up to $\Delta/\delta_{th}\approx 1$, the value of $f$ is near unity as one would expect, however, it quickly decreases for larger $\Delta$ values, indicating that the presumed PDF overestimates the filtered reaction rate for $\Delta/\delta_{th} > 1$. Note that the analytical expression of \eqref{eq:fdelta} requires $f(\Delta) \rightarrow 1$ for $\Delta \rightarrow 0$. A small error (less than 0.5\%) is however present in Fig.~\ref{fig:fdelta_propane} due to numerical integration and filtering of the numerator in Eq.~\eqref{eq:fdelta}, which is negligible for the purposes of the analysis conducted here.   

According to the above description, the following steps are used to impose the correction in the LES. First, the correction factor $f(\Delta)$ defined in Eq.~\eqref{eq:fdelta} is computed for the 1D freely propagating laminar flame for different values of $\Delta$ and pre-tabulated. 
The local filtered reaction rate integral over the cell volume appearing in the $\widetilde{c}$-equation in the finite volume approach is then pre-multiplied by $f(\Delta)$ in the LES at each time step according to the local mesh information, which 
enables the consumption speed correction. 
It is worth noting that, given its definition, the local value needed to compute $f$ should be in the flame-normal direction, and using $\Delta = \mathcal{V}^{1/3}$ is thus an approximation. This approach is here used for simplicity as this is the first attempt of using the proposed correction in an actual \textit{a posteriori} context. Moreover, for the unstructured grids used in this work the mesh elements are about equilateral and thus this error is neglected. However, this is to be taken into consideration in case of mesh elements with strong aspect ratio.

\section{Case study}\label{sec:CaseStudy}
The configuration chosen to \textit{a posteriori} validate the speed correction model described in Section~\ref{sec:Modelling_FlameSpeed} is the quasi-laminar jet flame studied experimentally in~\cite{FurukawaYW2016}. This configuration lies in the wrinkled flamelet regime of the Borghi diagram~\citep{Borghi1990}, see Fig.~\ref{fig:Borghi}, where the $\beta$-PDF is expected to overestimate the reaction rate. This configuration and its numerical modelling are described in this section. 

\subsection{Experimental details}
\begin{figure}[tb]
   	\centering
	\includegraphics[width=0.5\columnwidth]{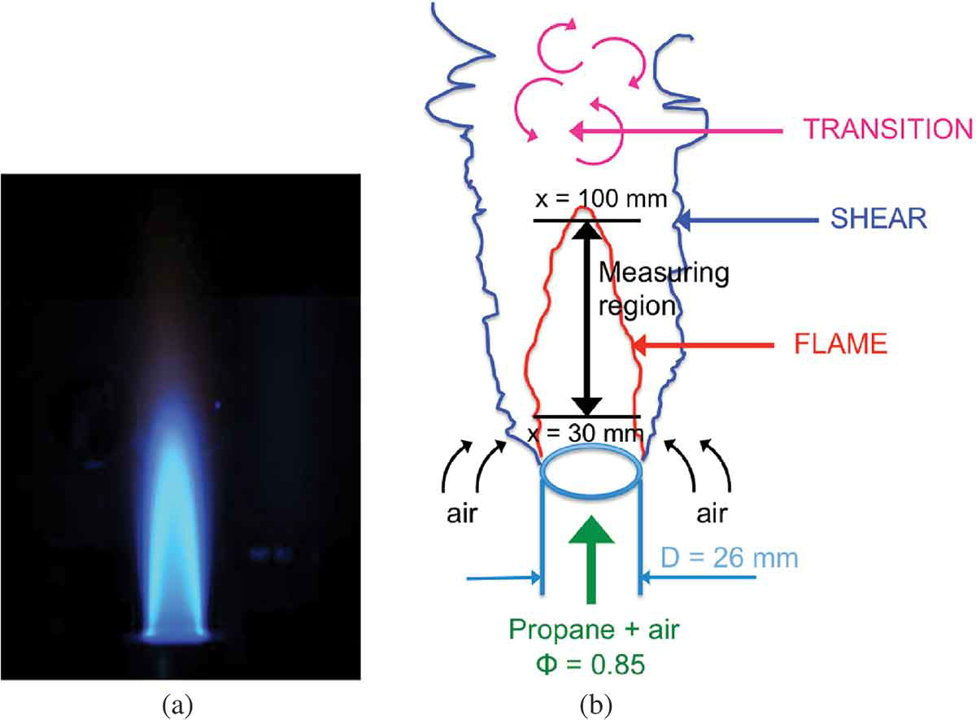}
	\caption{Direct photograph of the premixed jet flame (a) and its sketch representation showing relative position of flame and shear regions (b). Figure taken from~\cite{LangellaSWF2016}.}	\label{fig:WFlame}
\end{figure}

The experimental configuration consists of a premixed propane/air mixture issuing into a quiescent ambient at equivalence ratio $\phi=0.85$. This flame is part of the experimental campaign reported in~\cite{FurukawaYW2016}. A direct photograph and a schematic of the flame are shown in Fig.~\ref{fig:WFlame}.
%
The nozzle diameter is $D=26\,{\rm mm}$ and the bulk velocity in the cold flow mixture at temperature $T=298\,{\rm K}$ is $U_b=4\, {\rm m/s}$. As pointed out in~\cite{LangellaSWF2016}, the heat transfer between flame, wall and the fluid near the nozzle exit in the ignited configuration makes the mean centreline velocity at the nozzle exit to increase from $4.5$ to about $5.5\,{\rm m/s}$. Moreover, the absence of a pilot and the relatively low speeds result in a flame sitting within the jet central region and not in the shear layer, see Fig.~\ref{fig:WFlame}b. Consequently, the flame has a quasi-laminar behaviour due to the effect of inlet turbulence and self-induced fluctuations~\citep{LangellaSWF2016}. The mixture near the flame also does not mix with air in this configuration and thus remains premixed, except for the anchoring point.
Velocity measurements for this flame were taken with 3D laser Doppler velocimetry along the centerline and along the radius for given heights from the nozzle exit ranging between 30 mm and 90 mm. These measurement regions did not include the shear layer, i.e. they were focused in regions where shear-generated turbulence effects are weak, thus offering an appealing opportunity to validate modelling in quasi-laminar conditions.

\subsection{Numerical details and meshes}
\begin{figure}[tb]
   \centering
    
    \includegraphics[width=\columnwidth]{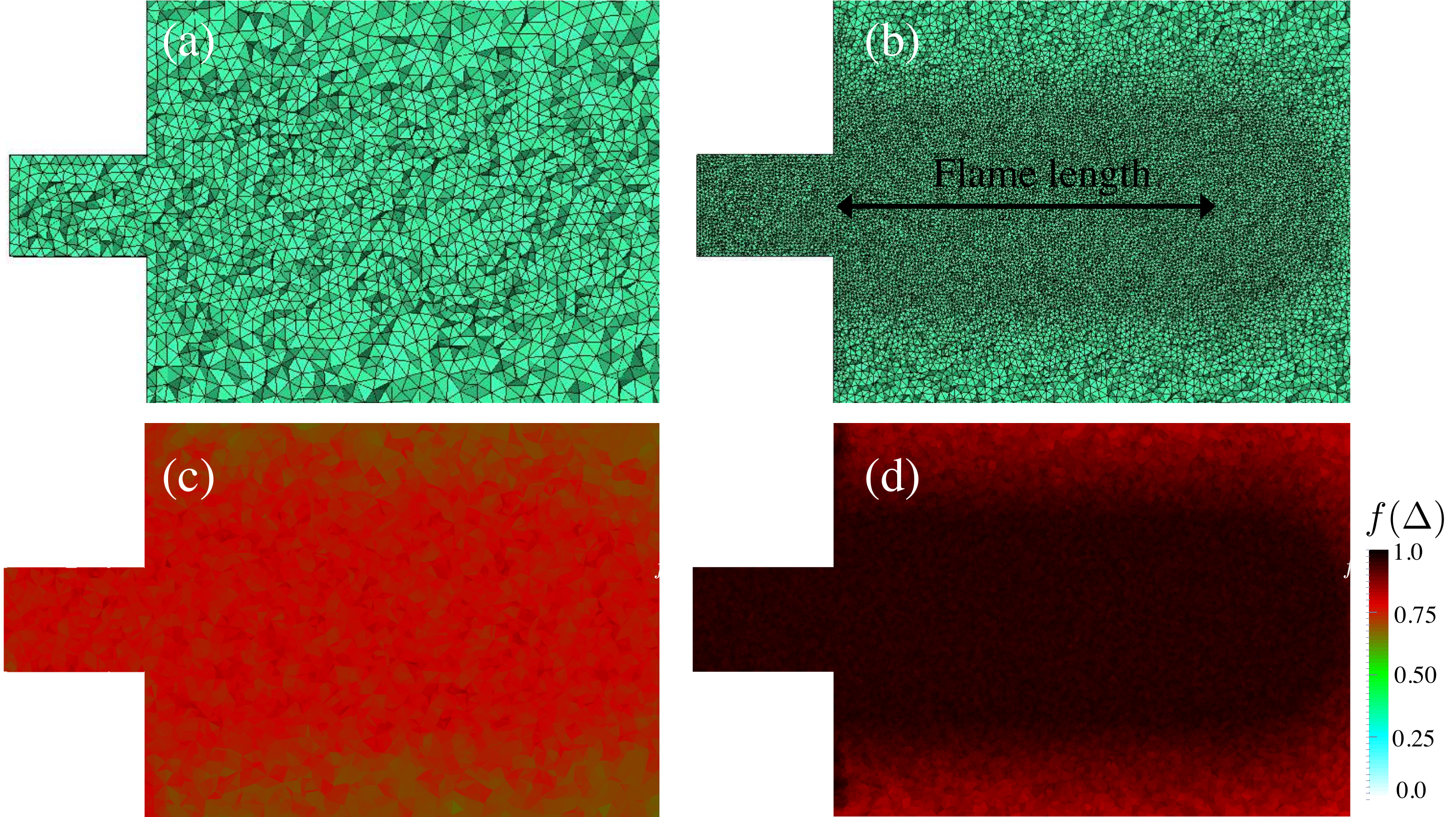}
	\caption{Zoom of the 1M (a) and 6M (b) meshes used for the premixed configuration near the nozzle exit; and corresponding contours of $f(\Delta)$ (c, d).}\label{fig:WMeshes}
\end{figure}

The numerical domain consists of the last $1.5D$ of the nozzle followed by a cylindrical region of $1\, \rm{m}\, \times \,1\,{\rm m}$. This domain is large enough to avoid potential numerical effects from the boundary, and is discretised using two non-structured meshes of 1M and 6M tetrahedral elements respectively, with refinement near the walls and the region of the flame, as shown in Figure~\ref{fig:WMeshes}.
    
%
These meshes consist of about 15 and 50 elements along the nozzle diameter respectively for the 1M and 6M meshes.
Values of $\Delta/\delta_{th}\approx 1$ and 3 are obtained for the 6M and 1M meshes in the flame region, respectively. These correspond to values of about $f(\Delta)\approx 1$ and 0.75 respectively, as shown in Fig. \ref{fig:fdelta_propane}.
The two meshes are used to analyse the effect of the consumption speed correction factor discussed in Section \ref{sec:Modelling_FlameSpeed}. As $f(\Delta)\approx 1$ in the fine mesh, the corresponding simulation provides a reference solution for the analysis to be discussed next. The coarse mesh will be used with and without the correction factor to assess the impact of that correction factor. This will thus allow to deduce whether the correction factor can improve the accuracy of an initial inaccurate simulation, at least in terms of flame speed and location.

The passive tracker, used to track the coflow, is assigned respectively values 1 and 0 in the main jet and air coflow boundary instead.
The progress variable is assigned values 0 and 1 for the reactant jet and entrainment inlet boundaries, respectively, while its variance is always zero on the boundary. 
Note that due to the presence of entrainment, the value of progress variable is not well defined at the coflow inlet in the premixed flame model. The choice of assigning this value to 1 does not imply that burnt gases are present since the passive tracker is zero here; on contrary, this choice is optimal for this configuration as it avoids the possible formation of a spurious flame in the shear layer (the progress variable would exhibit a gradient otherwise since the region between flame and shear layer is filled by burnt gases, see Fig.~\ref{fig:WFlame}b).

A temperature of 298 K and a flat velocity profile of 0.1 m/s are assigned to the coflow boundary to mimic the effect of entrainment. Slip-flow and zero-gradient conditions are assigned on the lateral boundaries for velocity and scalars respectively, except for the nozzle wall where a two-layer wall function model~\citep{PiomelliB2002} is used for velocity. 
A zero-gradient condition is applied at the outlet for all variables except pressure, which is given the atmospheric value. A turbulent velocity profile is assigned at the jet inlet, and the synthetic eddy method
described by~\cite{JarrinBLP2006} is used to mimic the effect of turbulent fluctuations with rms velocity $u'_{\rm rms} = 0.24\,{\rm m/s}$ and longitudinal and later integral scales respectively of $\Lambda_x = 10.9\,{\rm mm}$ and $\Lambda_y = \Lambda_z = 7.5\,{\rm mm}$~\citep{FurukawaYW2016,LangellaSWF2016}.
The turbulent velocity profile
 accounts for the effect of heat transfer at the wall as computed and discussed by~\cite{LangellaSWF2016}. A temperature profile is also used for the jet inlet accordingly. Furthermore, the temperature at the nozzle wall is assigned accordingly to the increase from 298 K to 403 K, 35 mm ahead of the nozzle exit observed in~\cite{LangellaSWF2016}. 
This way the heat losses to the wall are taken into account by means of wall temperature boundary conditions as done for example in~\citep{MercierAMDGVF2014,BenardLMM2019}. Nevertheless, this approach does not take into account possible heat loss effect on the reaction rate itself near the anchoring point. Although different methods for this have been developed in the context of flamelet methods (see for example \cite{FiorinaBGTCD2003,VanOijenG2000,ProchK2015,DoniniBOG2017}), 
modelling of heat losses is not considered in this study for simplicity, as it is unnecessary for the specific analysis on relative behaviour of different meshes discussed in this paper. Inclusion of heat loss effects is however postponed to a future work as it may significantly improve the LES prediction and further help understanding the physics near the anchoring point.
It is worth noting that satisfying comparisons with experimental data were presented in~\cite{LangellaSWF2016} by using similar assumptions, suggesting that statistics downstream the anchoring point can still be reasonably well predicted in this configuration despite heat losses being neglected in the thermochemistry table.

Simulations were performed on the Athena East Midland+ UK cluster using 196 cores in parallel. The time step for the LES is $\Delta t = 2\,{\rm \mu s}$ which guarantees a maximum Courant number below 0.2 on the 6M mesh. The time step is not increased on the coarse grid because a higher time step was observed to lead to numerical instability, whose causes are currently unclear. 
Statistics have been collected for a period of at least 6 flow-through times, $\tau_f$, after the statistically steady state is achieved, where $\tau_f$ is defined as the time a particle along the centreline of the jet takes at the bulk speed to move 100 mm downstream of the nozzle exit, since this distance is representative of the flame length. 
Each simulation took about 2 days on a wall clock to simulate a period of $6\tau_f$. Only about 2.5 hours are necessary on the coarser mesh to compute the same physical time.

\section{Premixed flame results}\label{sec:Results}

\subsection{Estimation of sub-grid variance}\label{Results_variance}
\begin{figure}[tb]
   \centering
    \includegraphics[width=0.5\columnwidth]{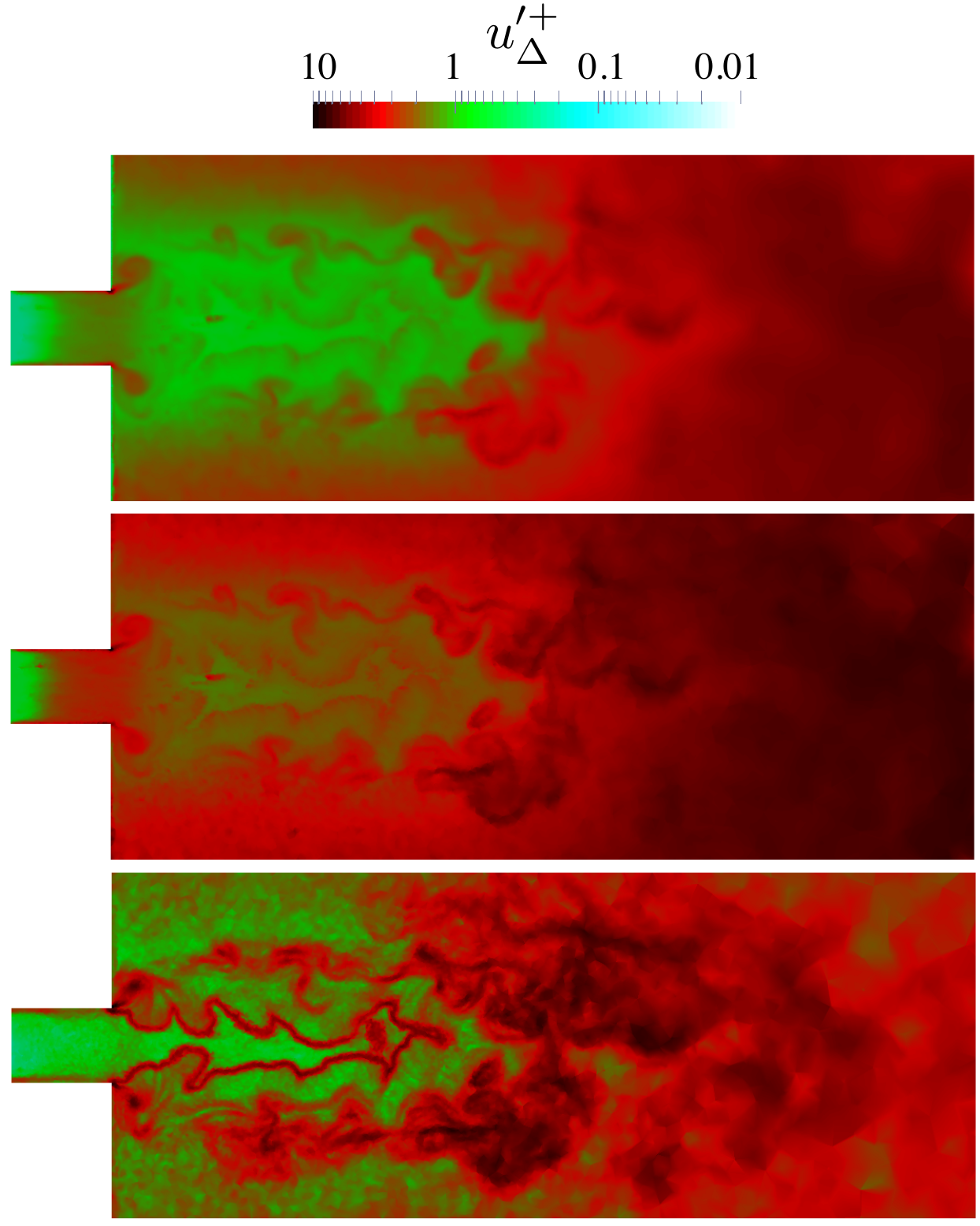}
     \caption{Midplane contours of normalised SGS velocity scale, $u^{\prime +}_\Delta = u^{\prime}_\Delta/s_L$ before and after the numerical flashback. The subgrid velocity \RA{is} obtained from subgrid kinetic energy transport equation (top), Lilly's model~\cite{Lilly1967} (middle) and the localised-dissipation model~\cite{LangellaDSP2018} (bottom).}\label{fig:FGM_uDelta}
\end{figure}

As discussed in Section~\ref{sec:CaseStudy}, the studied flame has a quasi-laminar behaviour and does not sit within the shear layer. This poses challenges for numerical modelling because any incorrect prediction of reaction rate and flame speed will immediately result in the flame moving upstream or downstream, making the chosen configuration particularly suited for the evaluation of the consumption speed correction. As discussed in the Introduction section, the flame consumption speed in presumed-PDF approaches can be overestimated for $\Delta/\delta_{th}>1$ even when the reaction rate is estimated correctly. On the other hand, an incorrect reaction rate also leads to an incorrect consumption speed. Since $\overline{\dot{\omega}}$ depends on $\sigma^2_c$ \RB{in the present modelling framework}, which in turn depends on $\widetilde{\varepsilon}_c$, it is important that the latter is correctly estimated \RB{in order to have a meaningful SGS variance}. Therefore,
as discussed in Section~\ref{sec:Modelling_combustion}, model C is preferred among those of Table~\ref{tab:SDR} as its underlying assumptions are better suited for the present quasi-laminar configuration. 
In order to find a value for the model constant $\beta_c$ in Eq.~\eqref{eq:SDRmodel}, a closure for the sub-grid kinetic energy $k$ has to be provided first, or equivalently a value for the subgrid velocity scale $u'_\Delta = \sqrt{2k/3}$. Thus, this quantity is investigated first. One must notice that $k$ is by definition a SGS kinetic energy and not a turbulent kinetic energy, thus its modelling has to capture both subgrid turbulent fluctuations and those produced by the flame pressure dilatation~\citep{LangellaDSP2018}. The simplest idea is to use directly $k$ from the transport equation in the LES. Nevertheless, this equation does not take into account the pressure dilatation terms as mentioned in Section~\ref{sec:Modelling_LES}. Thus, only the contribution of self-induced turbulence at the SGS scales is expected to be captured. A second method is to use the model from Lilly, $u'_\Delta = \mu_t/(C_L \overline{\rho}\Delta)$, where $C_L\approx 0.11$. Since $\mu_t$ depends on the velocity strain, this model is able to capture the effect of thermal expansion on the flame~\citep{ColinDVP2000}. However, \cite{LangellaDSP2018} pointed out that  the rotational contribution is also important, and proposed a different model based on localised dissipation (LD model):
\begin{equation}\label{eq:LDmodel}
  u'_\Delta = C_{LD} \left| \Delta^2 \nabla \widetilde{\mathbf{U}} : \nabla \widetilde{\mathbf{U}}  - \frac{1}{4} \left| \Delta^2 \nabla^2 \widetilde{\mathbf{U}} \right|^2 \right|^{1/2}
\end{equation}
where $\widetilde{\mathbf{U}}$ is the filtered velocity vector and $C_{LD}\approx 0.5$ for the conditions of the current flame~\citep{LangellaDSP2018}. Note that the diffusion term in the above equation is usually much smaller than the dissipation term and can be neglected.
The three different models are compared \textit{a posteriori} at a random time $t_0$ for the configuration of Fig.~\ref{fig:FGM_uDelta}, since the stronger reaction rate more clearly marks the effects on  $u'_{\Delta}$, which grows proportionally.
Note that the scale-similarity based model showing good performance in~\cite{LangellaSWF2016} is not a good candidate for this investigation because of the known issues of applying scale-similarity algorithms to unstructured meshes~\citep{VolpianiSV2016}.
%
The qualitative behaviors of the three models in the nearly-isothermal shear layer and the transitioned region downstream the flame tip are similar for all three models, and all predictions are of the same order of magnitude. The orders of magnitude within the flame, however, are very different. Only the SGS self-induced oscillations are captured when using $k$, which are of the order of the laminar flame speed, $s_L=0.34\,{\rm m/s}$. This is consistent with Fig.~\ref{fig:Borghi} where the flame lies at the boundary between corrugated and wrinkled flamelet regimes of the Borghi diagram. However, the amount of wrinkling observed in the figure suggests $u'_\Delta> s_L$, and this seems to be properly addressed only by the LD model.  This characteristic is to be kept in mind when identifying the flame regime in the Borghi diagram from the LES data. It is worth noting that the values of $u'_\Delta$ as predicted by the LD model for the stable flame configuration to be investigated in the next section are much smaller than those observed here, being of the order of $0.6$ m/s.

According to the analysis above, the model proposed by~\cite{LangellaDSP2018} (LD model) is preferred. Using model C of Table~\ref{tab:SDR} and Eq.~\eqref{eq:LDmodel}, a stable flame is found for a value of $\beta_c \approx 25$ in Eq.~\eqref{eq:SDRmodel}. This value provides
the correct flame length, which can be inferred from the peak positions of mean axial velocity and rms radial velocity along the centreline, to be discussed in Section~\ref{sec:Results_statistics}. 
Note that the sensitivity to this value was found to be relatively strong (a 10\% decrease of $\beta_c$ corresponds to $7.5\%$ shortening of the flame), which may be a consequence of the quasi-laminar configuration ($\mathcal{R}\gg \mathcal{P}$), and has to be kept in mind when model C of Table~\ref{tab:SDR} is used.

\subsection{General flame behaviour}\label{sec:Results_statistics}
According to experimental results~\citep{FurukawaYW2016}, the flame length $L_f$, defined as the distance of the flame tip from the nozzle exit, is about 100 mm, which can be inferred from the peak location of the centreline axial velocity. 
%
This is shown for the 6M grid in Fig.~\ref{fig:Comparisons}, indicating a good match between LES prediction using the 6M grid and experimental data for both axial velocity and radial rms velocity.
Since incorrect flame speeds would lead to incorrect flame lengths, the latter is used here as an indicator of the correct flame speed. In this respect, the computed flame on this grid will be used as the base reference flame for the analysis to be conducted in Section~\ref{sec:Results_2}. 

Additional velocity measurements were taken in~\cite{FurukawaYW2016}, which showed the  occurrence of a bimodal behaviour in the radial component of the velocity field across the flame at certain axial locations. Moreover, the velocity measurements were processed in the same work to construct progress variable isolines and flame brush thickness, which indicated the presence of a `bubble' near the flame tip. The further analysis of these statistics in the LES conducted by~\cite{LangellaSWF2016} indicated that the LES modelling with a presumed PDF approach was not capable of predicting such behaviours even with a mesh of 20 million elements. It was argued that for the velocity field to exhibit bimodality, the flame has to quickly move around the probing point. Consequently, the LES in~\cite{LangellaSWF2016} did not predict it either because the amplitude and speed of the flame movement in the radial direction was not captured accurately, or because a much finer mesh was required. Also, no `bubble' was observed in the LES. 
A further investigation conducted on LES data in the current study, however, indicates the presence of vortex rings forming at the nozzle exit due to the boundary layer detachment and interaction with the flame near the anchoring point. These vortices bring vorticity that travels downstream in the quasi-laminar region before being suppressed by thermal dilatation effects and it is possible that they affect the flame movement in the radial direction, which is something that could not be studied in~\cite{LangellaSWF2016} where the nozzle was not within the numerical domain.  
Nevertheless, since these ring vortices also cause some mixing with air in the region immediately near the flame anchoring point, their study in the context of the fully premixed model under investigation here would not be fully appropriate. Also,
although such an investigation would certainly be of interest, it goes beyond the purpose of this work, where only a reference flame is needed for the fully premixed, unstrained flamelet model. It is worth noting that, due to the reasons above, the use of a fully premixed model would not fully capture the vortex-flame interplay near the anchoring point as the flame is not fully premixed in that region. As this behaviour is linked to that of the radial velocity profiles, further experimental data available in~\cite{FurukawaYW2016} are not discussed here and this is left for future work, where the physics near the anchoring point will be better investigated also in the context of the possible differential diffusion and heat losses effects discussed earlier.
Nevertheless, this is not relevant for the analysis in the next section, namely demonstrating the effectiveness of the flame consumption speed correction discussed in Section \ref{sec:Modelling_FlameSpeed}. Indeed, the spurious increase of consumption speed can be verified just by the comparison of the numerical simulations on different grids.
The flame investigated in this section will thus be used as a reference (baseline case)  for the analysis in the next section.

\subsection{Behaviour on different meshes and flame speed correction}\label{sec:Results_2}

\begin{figure}[tb]
   \centering
    \includegraphics[width=0.8\columnwidth]{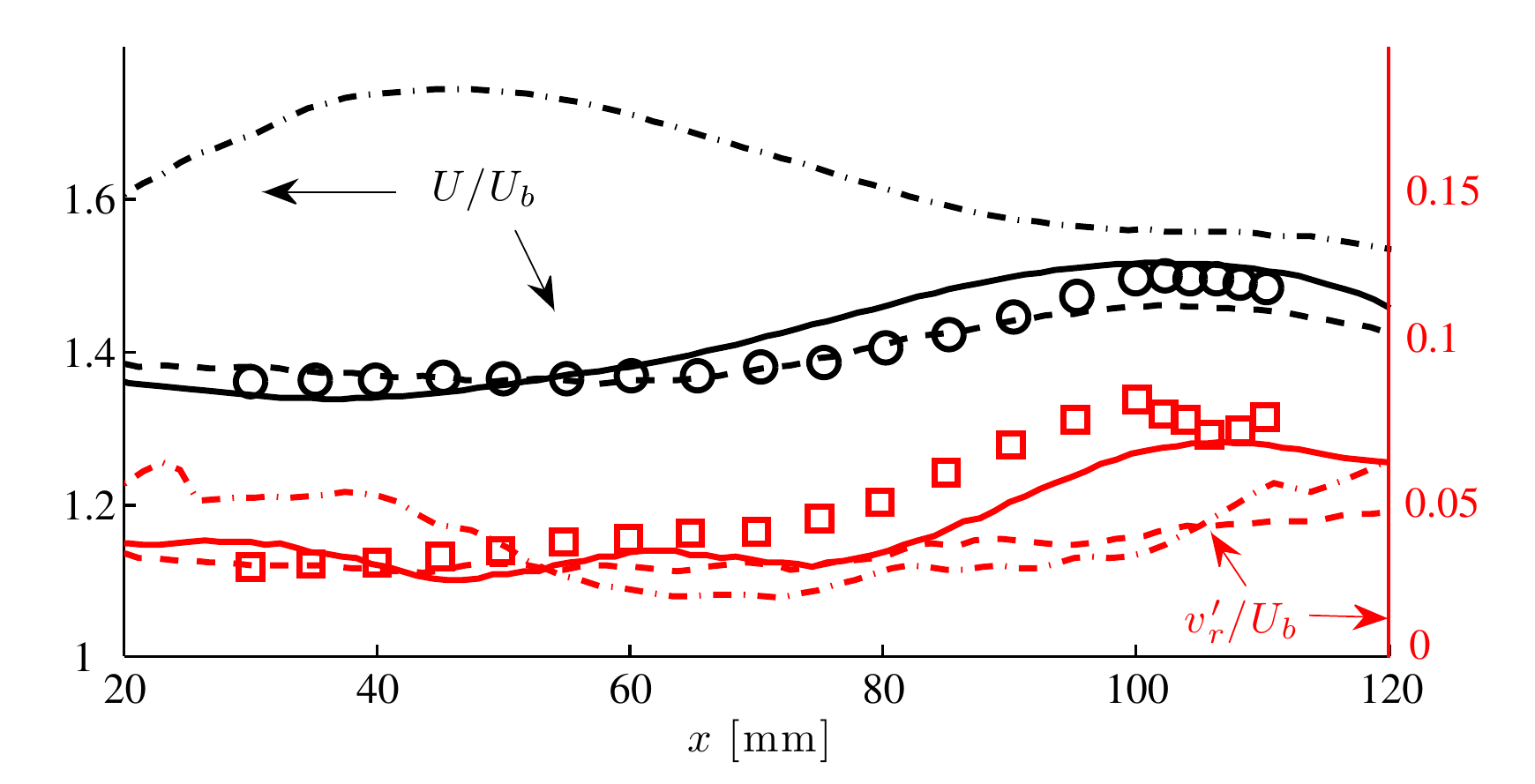}
     \caption{Centreline variation of mean normalised axial velocity, $\langle U \rangle/U_b$ and radial rms velocity, $\sqrt{\langle v'_r \rangle}/U_b$ obtained using model C of Table~\ref{tab:SDR} are compared with experimental data~\cite{FurukawaYW2016} (symbols) for the 6M ($\ldash{black}$) and 1M meshes with ($\mdash{black}\;\mdash{black}$) and without ($\mdash{black}\;\sdash{black}\;\mdash{black}$) the flame speed corrector of Eq.~\eqref{eq:CorrectedOmega}. Values for $U/U_b$ for the 1M grid without flame speed corrector are multiplied by a factor $a=0.75$ to fit to the scale.}\label{fig:Comparisons}
\end{figure}

\begin{figure}[tb]
   \centering
\subfloat[]{\includegraphics[width=0.5\columnwidth]{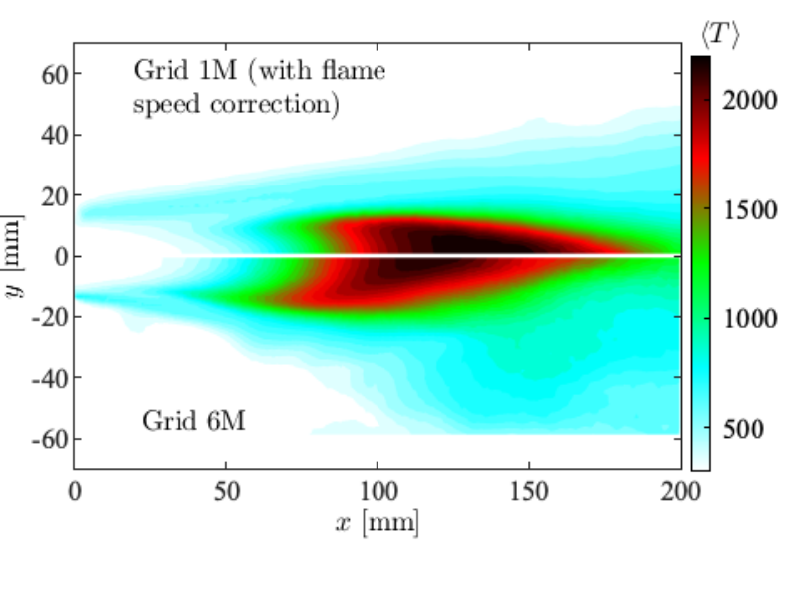}}
\subfloat[]{\includegraphics[width=0.5\columnwidth]{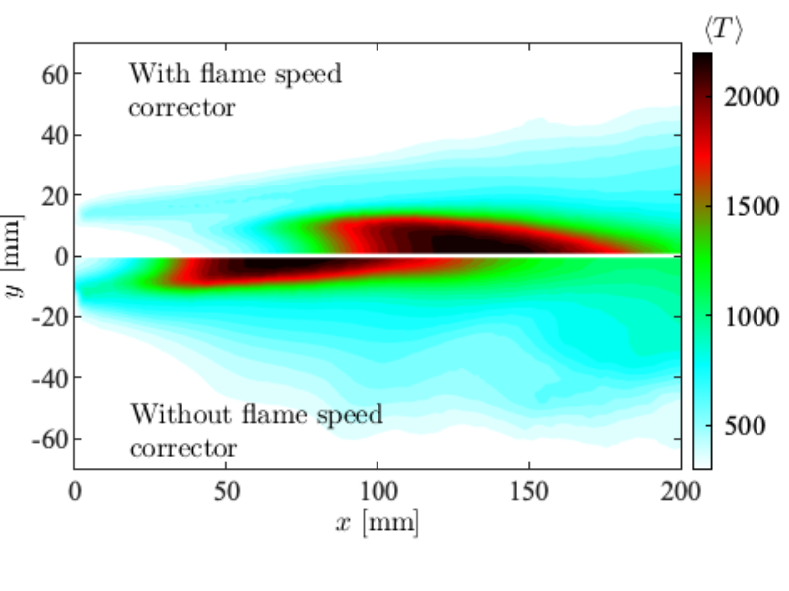}}\\
\subfloat[]{\includegraphics[width=0.5\columnwidth]{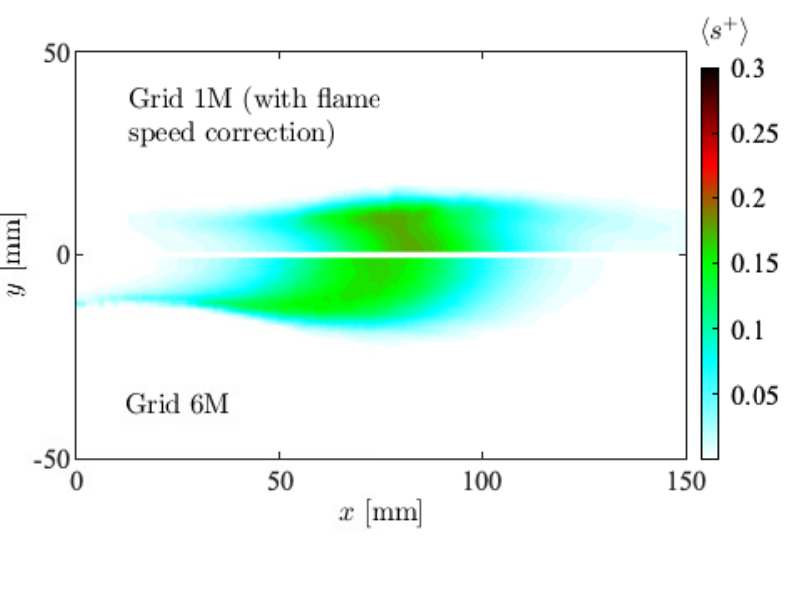}}
\subfloat[]{\includegraphics[width=0.5\columnwidth]{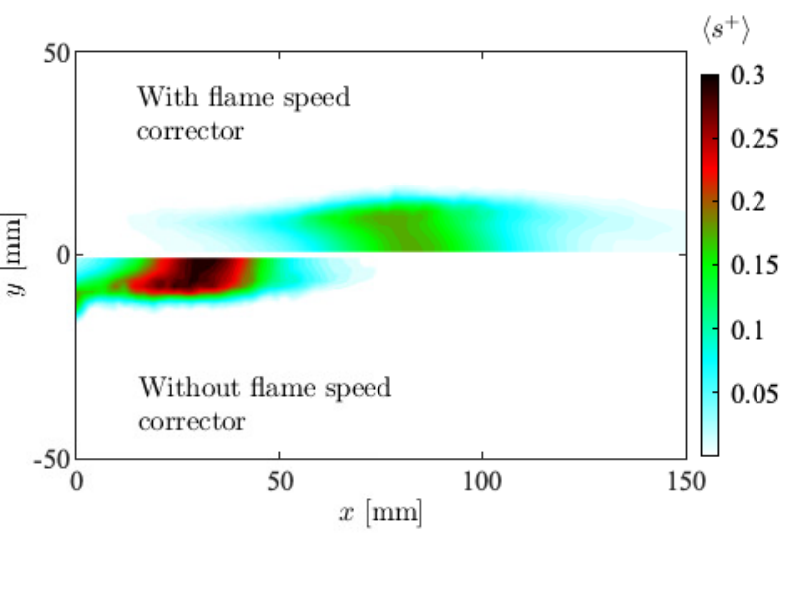}}
\caption{Contours of mean temperature (top) and mean normalised \textcolor{Green2}{flame} speed $s^+ = f(\Delta)\overline{\dot{\omega}}_c \delta_{th}/(\rho^b s_L)$ (bottom) obtained from LES using the 6M grid and the 1M grid with and without flame speed correction.}\label{fig:MeanContours}
\end{figure}

When using the 6M grid, model C of Table~\ref{tab:SDR} with $\beta_c\approx 20$ provides the correct amount of subgrid SDR on the flame to achieve the correct flame length as compared to experiment, and consequently the correct flame speed. This is observed in Fig.~\ref{fig:Comparisons}, where centreline variations of mean axial velocity and radial rms velocity from LES are compared to those from experiments~\cite{FurukawaYW2016}. 
The velocity increase near $x\approx 100\,{\rm mm}$ is in fact caused by the thermal dilatation from the flame tip. Further improvements could be achieved by adjusting the combustion constants and the inlet turbulence. However, this goes beyond the purpose of this work and for the same reason additional comparison with experimental data is not shown. The objective of this section is instead to assess whether, for the identical numerical setup used, the same flame length is retained using the 1M grid. As expected, when switching to the 1M mesh, the flame accelerates due to overprediction of the reaction rate, as shown in Fig. \ref{fig:Comparisons}. The flame length, indicated by the new position of the axial velocity peak at $x\approx 45\,{\rm mm}$, decreases  by a factor of  2 and the increased heat release causes the peak velocity to increase by a factor 1.5. The strong effect of flame dilatation is also reflected in the values of the radial rms velocity, that are observed to increase near the flame tip. This new flame length and position are statistically steady. The flame speed corrector, Eq.~\eqref{eq:CorrectedOmega}, is then switched on at this point. Contours of mean temperature are shown for the two grids in Fig.~\ref{fig:MeanContours}a and b. One can immediately notice that the flame speed correction is effective in switching the flame position back in the right position as compared to the 6M mesh results, and this is also reflected in the velocity comparisons of Fig.~\ref{fig:Comparisons}, where both peak position and values are now very similar to those obtained using the 6M mesh. Comparisons of radial velocity rms also improve for $x<80\,{\rm mm}$, however some underestimation is observed for downstream positions which could be caused by numerical diffusion due to the coarser mesh. 
One can expect that the artificial viscosity introduced by different mesh resolutions can also play an effect on the comparisons of Fig.~\ref{fig:MeanContours} and 
this effect could not be quantified or corrected in this work. This aspect has to be kept in mind when comparing quantities on different meshes. 

In order to demonstrate the effectiveness of the flame speed corrector, the consumption speed $s_c$ is compared for the 6M grid and the 1M grid with and without using Eq.~\eqref{eq:CorrectedOmega}. A local consumption speed in the current LES can be defined according to the product-based progress variable as
\begin{equation}
    s_c = \frac{1}{\Psi^b \rho^b}\int_\mathcal{V} \overline{\dot{\omega}}_c \;d\mathcal{V}
\end{equation}
where $\Psi^b = Y_{\rm CO_2}^b + Y_{\rm CO}^b$ and the superscript $b$ refers to burnt conditions.
 Given the consumption speed is an extensive quantity and is volume-dependent, in order to compare results for the two grids 
 a normalised reaction rate is derived by dividing $s_c$
 by the volume of the local numerical cell. Considering Eq. \eqref{eq:CorrectedOmega}, and since $\Psi_b$, $s_L$ and $\delta_{th}$ are constant in premixed flames, this normalised reaction rate 
can be constructed
 as $s^+ = f(\Delta)\overline{\dot{\omega}}_c \delta_{th}/(\rho^b s_L)$, where $f(\Delta)=1$ for the 6M grid and is either unity or taken from Eq.~\eqref{eq:fdelta} on the 1M grid depending on whether the flame speed corrector is used or not. Note that this quantity is meant here to represent a (normalised) flame speed rather than a reaction rate and thus is in principle,
 different from $\overline{\dot{\omega}}^+ = s^+\rho^b/(\overline{\rho}f)$ defined earlier,
 albeit the two mathematical expression are formally very similar.
Contours of time averaged $s^+$ are shown in Figs.~\ref{fig:MeanContours}c and d. As one can observe, a reasonable agreement between $s^+$ is obtained between
the 6M grid and the 1M grid when the correction $f(\Delta)$ is used, which explains why the flame remains in the same position. More in detail, the flame length seems to be quite well recovered, while some discrepancy can still be observed in terms of flame width.
On the contrary, values twice as strong are found for the 1M grid, which explains its positioning upstream where local reactants flow rates are higher. 
%

\begin{figure}[tb]
   \centering
\includegraphics[width=\columnwidth]{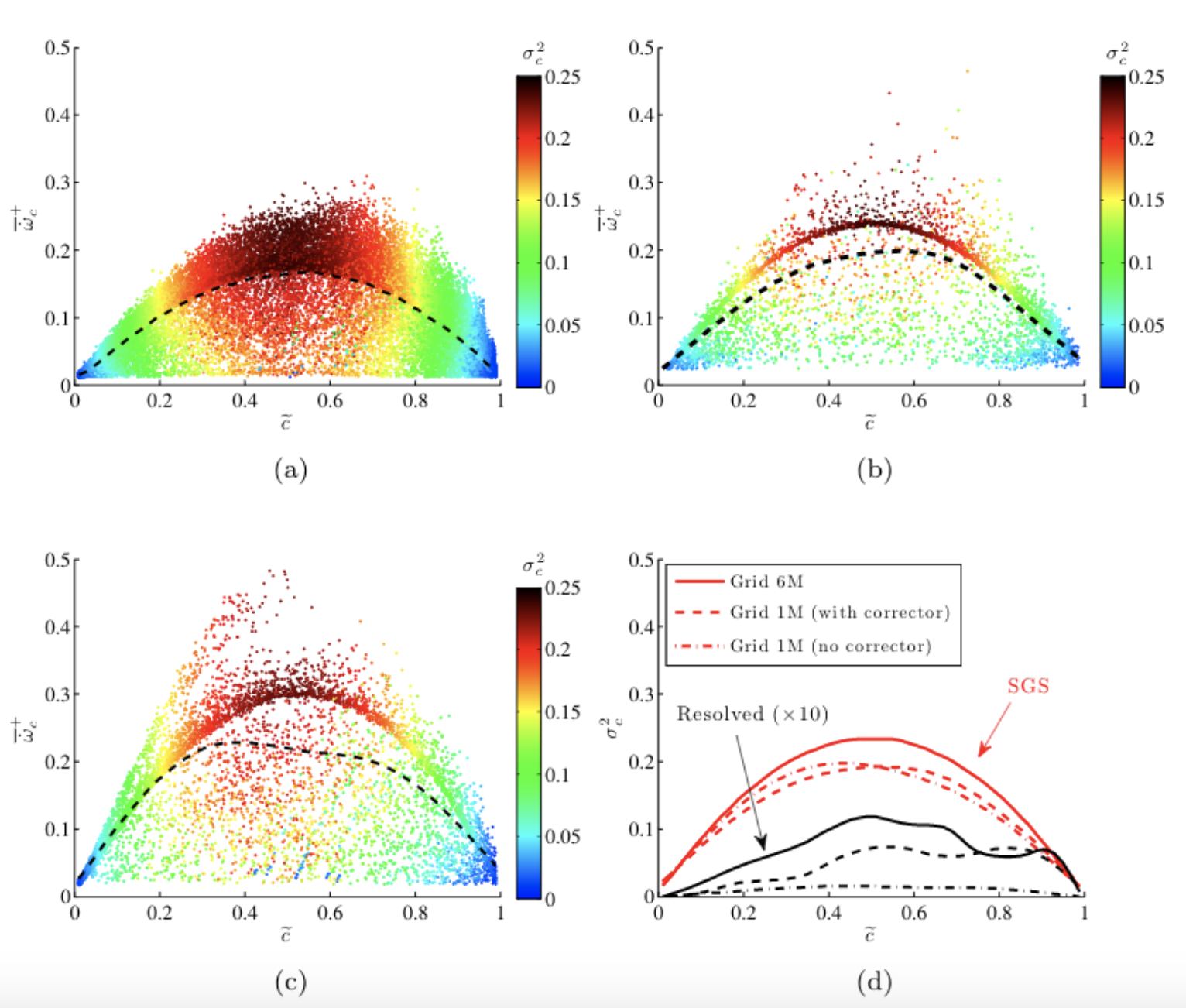}
\caption{Scatter plots of normalised reaction rate, $\overline{\dot{\omega}}_c^+ = \overline{\dot{\omega}}_c \delta_{th}/(\overline{\rho} s_L)$ for the 6M grid (a), and the 1M grid with (b) and without (c) flame speed correction. The plots are coloured by SGS variance and the dashed line indicates the conditional average. The conditional averages for subgrid and resolved variance of $\widetilde{c}$ are shown in (d).}\label{fig:OmegaScatters}
\end{figure}

\begin{figure}[tb]
   \centering
   \includegraphics[width=\columnwidth]{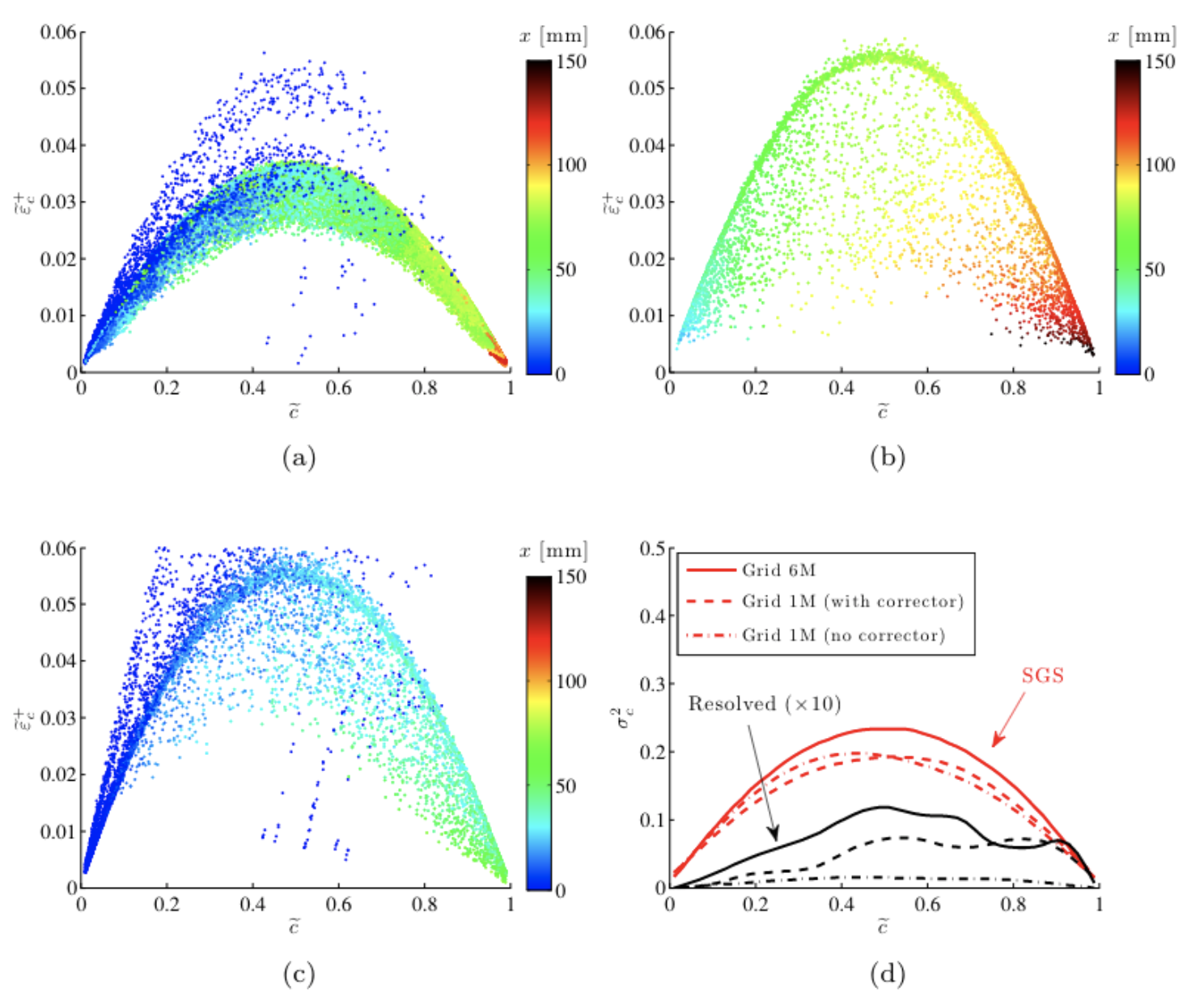}
\caption{Scatter plots of normalised subgrid SDR, $\widetilde{\varepsilon}_c^+ = \widetilde{\varepsilon}_c \delta_{th}/s_L$, for the 6M grid (a), and the 1M grid with (b) and without (c) flame speed correction. The plots are coloured by axial position. The conditional averages are also shown (d).}\label{fig:SDRScatters}
\end{figure}

Further insight is provided by looking at the scatter plots of Fig.~\ref{fig:OmegaScatters}, showing the behaviour of the normalised reaction rate conditioned on progress variable for the two grids.
%
These scatter plots help to understand the behaviour of the reaction rate as it is obtained from the flamelet table. For a certain value of $\widetilde{c}$, this depends on the value of SGS variance, which is shown in colours in the plots. 
First, one can observe that the values of conditional normalised reaction rate are larger for the 1M grid in the case without correction, which reflects the fact that the flame is positioned in a region of higher reactant velocities and thus higher reaction rates are required. It is worth noting that the mesh is nearly uniform in the region of the flame as can be observed from Fig.~\ref{fig:WMeshes}. Values of conditional SGS variance shown in Fig.~\ref{fig:OmegaScatters}d for the 1M grid cases with and without correction are also similar, suggesting that the higher reaction rates are a result of a decreased variance at the LES resolved level. In fact, the resolved variance is observed to decrease by a factor of 5, from about 0.01 in the correct flame position, to 0.002 in the shorter flame. Although both values are relatively low compared to the respective SGS values (because the combustion is at SGS level on these grids), this difference is still sufficient to yield the observed differences in the mean reaction rate behaviour. It is worth noting that the high values of SGS variances observed are a consequence of the strong reactive term in the variance equation and a relatively low scatter for fixed values of progress variable.
The behaviour of the subgrid SDR is investigated in Fig.~\ref{fig:SDRScatters}. As observed, the conditional values increase on the coarser grids, indicating that the modelled behaviour of the SDR is correct at least qualitatively. Moreover, the conditional behaviour does not change on the 1M grid between the cases with and without flame speed correction, where only a shift towards more upstream axial locations is observed to reflect the shorter flame.

The SGS variance behaviour observed for Fig.~\ref{fig:OmegaScatters}d can be counter-intuitive as lower values are observed for the coarser mesh. When the flame speed correction is not used, this is simply a consequence of the different flame topology (the flame stabilises in a different position). As the correction model only affects the burning speed and not the variance, however, the variance value does not come back to the values observed for the 6M grid. This poses questions on the meaning of the SGS variance, which is due to the fact that this variance is a controlling parameter in presumed PDF approaches. One could in fact equivalently increase the SGS variance magnitude by tuning the SDR model constant. One could also, in principle, estimate the $\Delta \sigma^2_c$ required to achieve the equivalent $\Delta \overline{\dot{\omega}}_c$ after applying the burning speed correction presented here. This is beyond the purposes of this work. 

%


\section{Generalised scaling for partially-premixed flames} 
  \subsection{Model description}\label{sec:PPflame_description}
  
\begin{figure}[htb]
   	\centering
    \subfloat[]{\includegraphics[width=0.45\columnwidth]{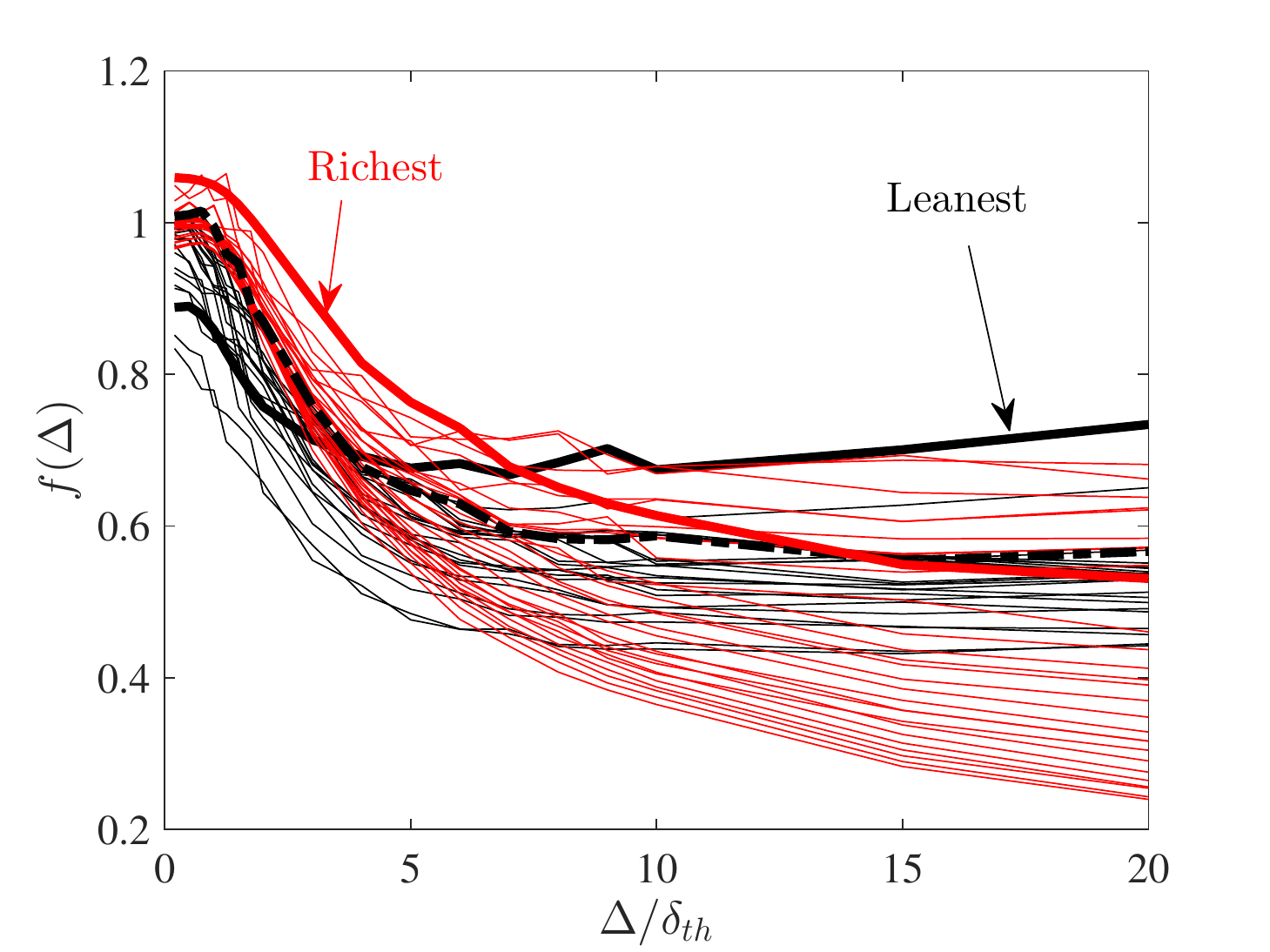}}
    \subfloat[]{\includegraphics[width=0.45\columnwidth]{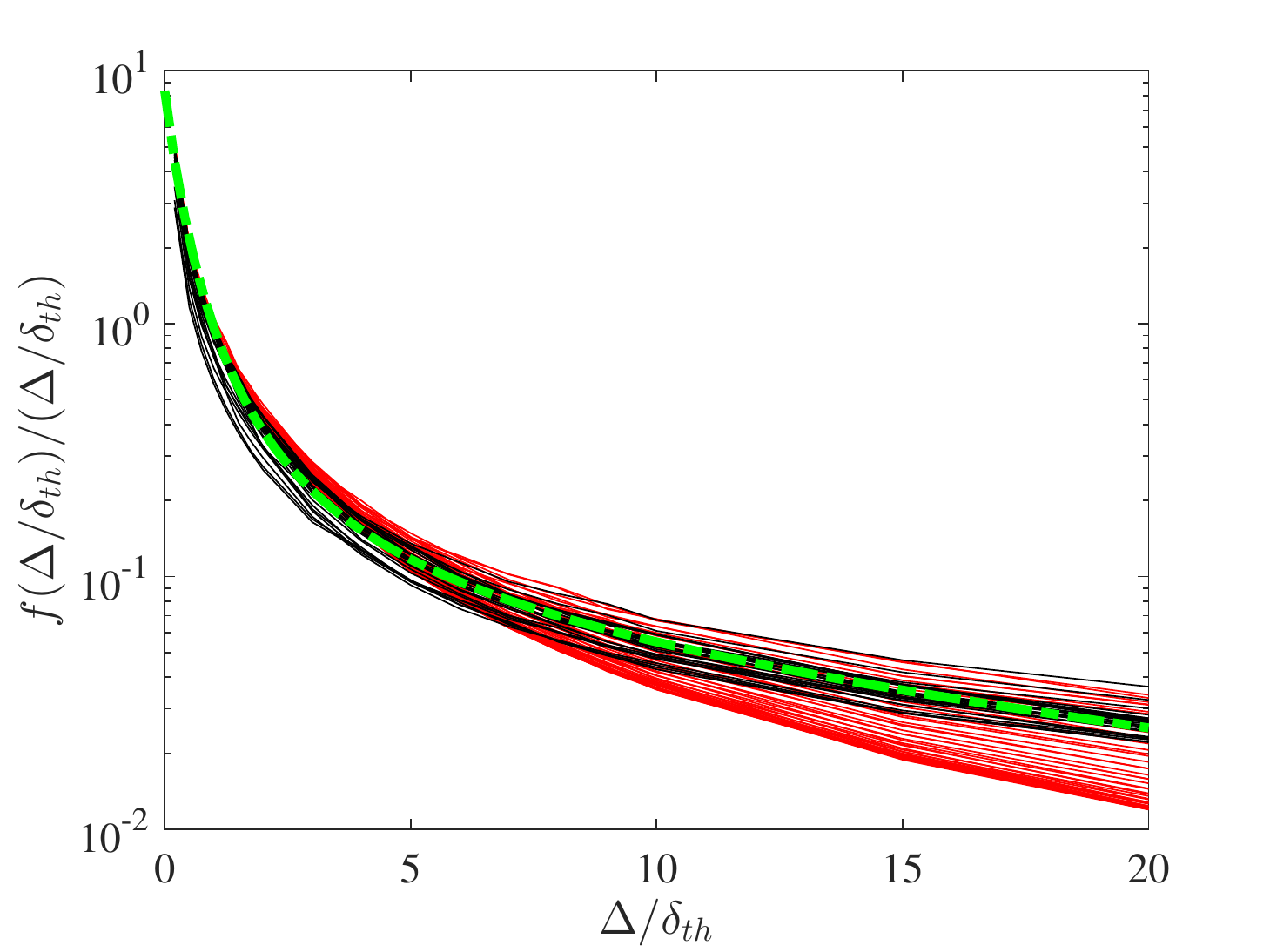}}\\
    \subfloat[]{\includegraphics[width=0.45\columnwidth]{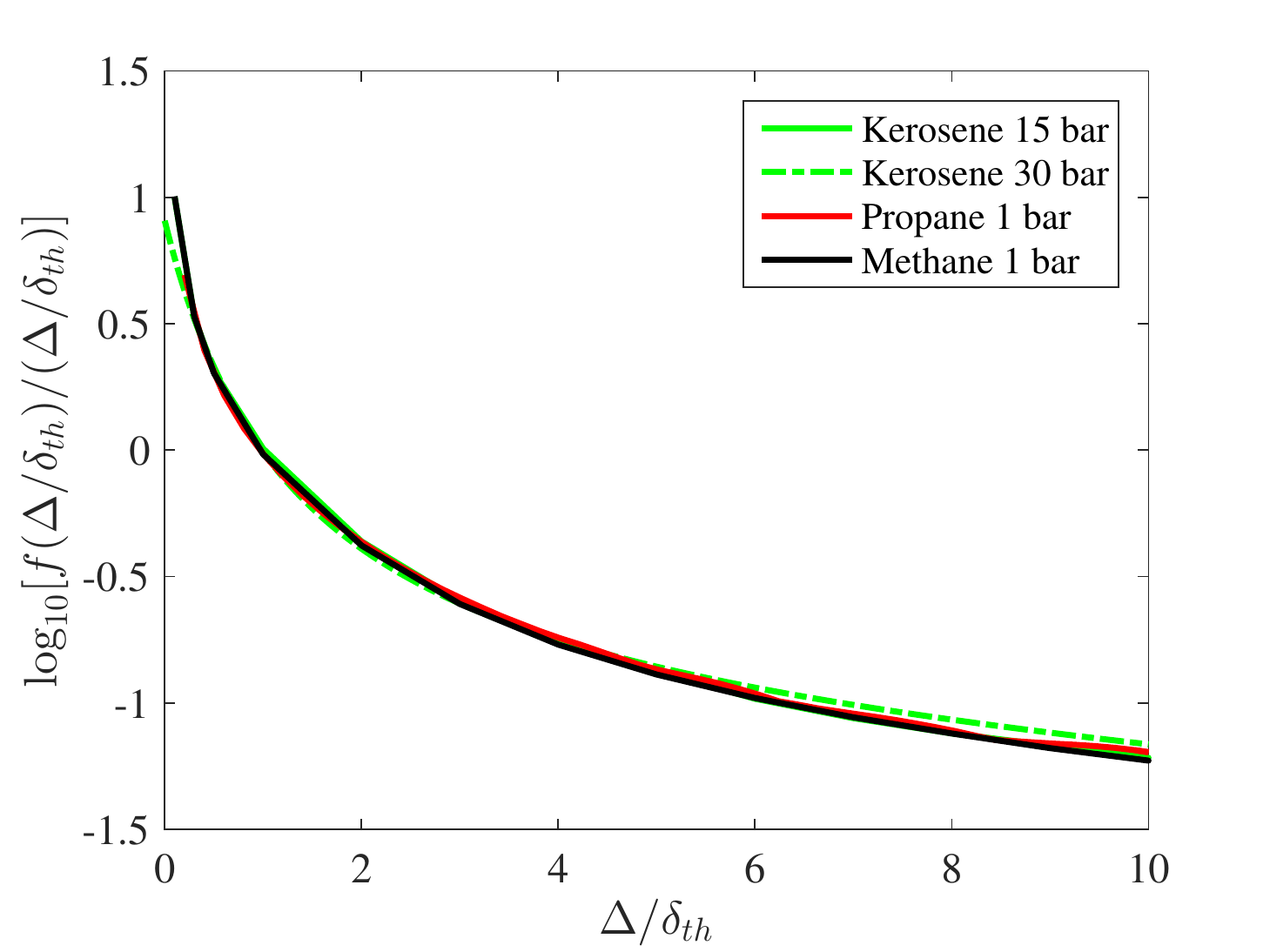}}
	\caption{Scaling function $f(\Delta/\delta_{th}$) for kerosene/air mixture at high pressure and different values of lean ($\ldash{black}$) and rich $\ldash{red}$ equivalence ratios spanning the flammability limits. The stoichiometric curve is emphasised ($\mdash{black}\;\sdash{black}\;\mdash{black}$). The same curves scaled by $(\Delta/\delta_{th})$ are shown in logarithmic scale in (b) along with the fitting curve ($\mdash{Green2}\;\sdash{Green2}\;\mdash{Green2}$). 
	The fitting curves for different fuels and pressures are shown in (c).
	}	\label{fig:fdeltas_kero}
\end{figure}

The flame speed corrector described in Section~\ref{sec:Modelling_FlameSpeed} can also be employed for partially premixed cases by computing $f(\Delta)$ for different flamelets at different equivalence ratios and storing the values in a precomputed table. For the high-pressure kerosene flame to be investigated in Section~\ref{sec:PPflame_testcase} the scaling functions are shown in Fig.~\ref{fig:fdeltas_kero}a.
%
It is worth noting that the values for $\delta_{th}$ are different for each flamelet, and thus the actual range of $\Delta$ is  different for each curve. However, the position of the peak does not change for $f(\Delta/\delta_{th})$, which is not the case for $f(\Delta)$. From this information and further scaling $f(\Delta/\delta_{th})$ by $(\Delta/\delta_{th})$ one obtains the graph in Fig.~\ref{fig:fdeltas_kero}b, which is shown in logarithmic scale for convenience. In this scale the curves seem to have a much better collapse, which allows to look for a fitting curve in this space. Note that the scaling with $(\Delta/\delta_{th})$ does not give significant advantages in terms of percentage variation between two curves at a fixed $\Delta/\delta_{th}$, but allows to achieve a monotonic variation. Moreover, the curves do not cross each other anymore if rich and lean flamelets are considered separately. These properties are convenient for the following analysis.

The next step is to find an analytical expression to fit the function $\log_{10}(f^+)$, where $f^+=f(\Delta/\delta_{th})/(\Delta/\delta_{th})$.
By inspection of the curves, the following scaling is suggested:
\begin{equation}\label{eq:rat21}
    \log_{10} f^+(x) = \frac{p_1 x^2 + p_2 x + p_3}{x+q_1}
\end{equation}
This scaling ensures the asymptotic convergence to $f(\Delta) = 1$ for $\Delta \rightarrow 0$, and that the coefficients are of a similar order of magnitude thus avoiding possible truncation errors in the LES.
The best fit is computed for the lean conditions only and found for $p_1=-0.0253$, $p_2=-1.373$, $p_3=1.354$ and $q_1 = 1.469$. The corresponding curve is shown in Fig.~\ref{fig:fdeltas_kero}b. It is also interesting to understand the behaviour of $f^+$ for different fuels and pressures, which is done in Fig.~\ref{fig:fdeltas_kero}c, where the fitting curves are computed for kerosene at 15 bar and 30 bar to represent real GT conditions, and for propane and methane at atmospheric condition. 
As can be observed from the figure, all curves are very similar suggesting that $f^+(\Delta/\delta_{th})$ does not strongly depend on fuel and pressure and thus the same fitting curve can be used for numerical modelling purposes in LES under different configurations.
 
\subsection{Test case: bi-stable combustor of aeronautical interest}\label{sec:PPflame_testcase}
\begin{figure}[tb]
   	\centering
	\includegraphics[width=0.7\columnwidth]{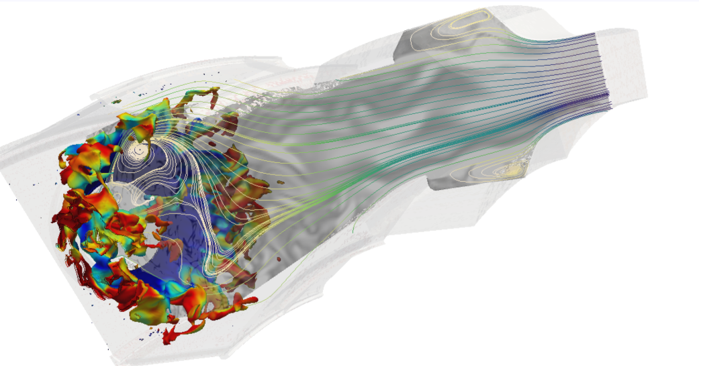}
	\caption{Sketch of one sector of the Rolls-Royce combustor Alecsys showing iso-contour of scaled progress variable $\widetilde{c}=0.5$ coloured by temperature (red-hotter, blue-colder), and centreplane mean streamlines on top of vorticity contour. Figure taken from~\cite{CORNET_URL}.}	\label{fig:RRgeom}
\end{figure}

The generalised flame speed corrector discussed in Section~\ref{sec:PPflame_description}, Eq.~\eqref{eq:rat21}, has been implemented in the Rolls-Royce in-house code Precise-UNS~\citep{AnandESZZ2013} to assess its performance under GT conditions. The Rolls-Royce lean-burn developmental combustor Alecsys is used for this assessment. Only one sector of the annular configuration is simulated here and a sketch is shown in Fig.~\ref{fig:RRgeom}.
%
Swirled, preheated air at about 800 K divides into a central pilot stream and a surrounding, main stream before mixing with the sprayed kerosene. A strong recirculation region is produced that entraps hot gases and provides the stabilization mechanism for a lifted flame. The bulk velocity at the inlet is about 180 m/s and the operating pressure above 30 bar, which is representative of high-power flight condition.
Effusion cooling arrangements and six film cooling slots are used at the combustor walls. The flame spans a range of different conditions between the thickened flame regime up to the distributed reaction zone regime of the Borghi diagram, as shown in Fig.~\ref{fig:Borghi}. 
The use of flamelet-based approaches for GT combustors is controversial in the classical viewpoint, however a number of works (see for example~\cite{DunnMBB2010,TemmeWSD2015}) have demonstrated that flamelet structures are still present at GT regimes. In fact, flamelets can be distributed over a wider region yielding a thicker flame brush rather than be thickened by turbulence. Moreover, small eddies may not have enough energy to impart significant changes to the flame internal structure~\citep{PoinsotVC1991,RobertsDDG1993,DoanSC2017}. Thus the limits of the flamelet assumption are not well defined. 

The flame in Alecsys is designed to be sufficiently away from the injector, and has a resulting M-shape configuration (the reader can refer to Fig. 1 of~\cite{LangellaHBVSZ2020} for a schematic). On the other hand, the intricate balance between heat release, turbulence and mixing in this configuration poses challenges for the numerical modeling and a more stable, V-shape configuration can be observed instead~\citep{LangellaHBVSZ2020}. 
The expected M-shape was predicted  by~\cite{SemlitschHLSD2019} for this configuration using model C of Table~\ref{tab:SDR} with a dynamic evaluation of the combustion model constant; however, whether this shape was a result of an optimised value of the model constant (via dynamic evaluation) in the SDR model was not investigated.
Additional simulations using the subgrid SDR models A and B of Table~\ref{tab:SDR} are carried out in this study and observed to lead respectively to a too-weak M-flame and a strong V-flame near the pilot, regardless of the value used for the model constant within the range indicated in Table~\ref{tab:SDR}. Since the underlying assumption of model B are better verified in this configuration ($\mathcal{P}$ and $\mathcal{R}$ in Eq.~\eqref{eq:cvar} are both of leading order, see discussion in Section~\ref{sec:Intro}),
although no experimental data is available for quantitative validation, a qualitative analysis is conducted in this section to assess if the reactive flow field predicted by this model (V-flame) is a result of an overestimation of consumption speed due to $\Delta \gg \delta_{th}$ and can thus be corrected by the flame speed correction described in Section~\ref{sec:PPflame_description}.
Note that the procedure followed for the premixed case of Section~\ref{sec:CaseStudy} cannot be pursued here as the tuning of the model constant has to be done for a mesh size with $\Delta \approx \delta_{th}$, which is beyond our computational reach.
It is also worth mentioning that the presence of spray adds additional uncertainty for the evaluation of the speed correction performance due to the possible interaction of long-lasting droplets with the flame~\citep{MaR2016}. Nevertheless, it was shown for a very similar injector in~\cite{LangellaHBVSZ2020} that most droplets evaporate before reaching the reacting region and so the aforementioned effect is possibly negligible here. On the other hand, the bi-stable nature of the chosen configuration is an important characteristic for the flame speed correction model since different burning speeds lead to different flame positions, so allowing for an assessment, albeit qualitative, which would otherwise not be possible without detailed experimental data.

  \subsubsection{Numerical details}
  The Rolls-Royce code Precise-UNS uses the finite volume approach and the SIMPLEC algorithm~\citep{VanDoormaalR1984} with 5 sub-iterations to discretise the equations described in Section~\ref{sec:Modelling}. Numerical discretisation schemes and limiters are the same as those described for OpenFOAM in Section~\ref{sec:Modelling_LES}, except for the time discretisation which is also second order and the use of relaxation factors of 0.4 and 0.7 for pressure and all other variables respectively.
  
  
  The premixed flamelets database for kerosene/air is computed using the mechanism of~\cite{DagautC2006} and by linearly varying the reactants temperature from the air condition to that of the evaporated fuel (about 650 K). The sprayed kerosene is modelled using a coupled Eulerian–Lagrangian approach without secondary Schmehl but reduced initial Sauter mean diameter compared to~\cite{SemlitschHLSD2019}, and Rosin–Rammler distribution. The droplets are injected at constant speed and random injection angle sampled using a Gaussian distribution with a 10 deg standard deviation and the mean direction aligned with the injector axis. A rapid mixing formulation is used to model the droplet evaporation, where the liquid kerosene is assumed to have very large thermal conductivity. 
  
  Boundary conditions are assigned as in~\cite{SemlitschHLSD2019} with inlet velocity profiles estimated from preliminary RANS simulations where the plenum is included in the numerical domain, and provided by Rolls-Royce. No inlet turbulence is provided as most of the turbulence is generated by the shear layers and recirculation regions. The mesh used for the simulation is an hexa-dominant unstructured mesh with about 11M elements, which is similar to that used in~\cite{SemlitschHLSD2019} and satisfies Pope's 80\% TKE rule except for the regions near the walls where the boundary layer is not resolved. The reader can refer to~\cite{SemlitschHLSD2019} for additional numerical details including typical computational times. Note that, given the relatively high computational cost of these simulations, statistics are not collected as they are not relevant to the analysis to be conducted here. 
  The mesh results in values of $\Delta/\delta_{th}$ between 1.6 and 30 in the region of the flame, which corresponds to values of $0.5<f(\Delta)<0.87$. These values are sufficient to trigger a change of flame configuration from V-shape to M-shape. Unfortunately the multi-grid analysis conducted in Section~\ref{sec:Results} cannot be repeated here because a computationally too-expensive mesh would be needed to obtain values of $f(\Delta)\approx 1$. 
 For the same reason the choice of model C for this type of configuration is less obvious because of the lack of generalities in the value to assign to the model constant $\beta_c$\footnote{The reader is reminded that dynamic evaluations of model constants are not attempted in this work for the reasons explained in Section~\ref{sec:Modelling_combustion}}. 
  On the contrary, the underlying assumptions of model B are now better verified (the turbulence production term $\mathcal{P}$ in the SGS variance equation is of leading order in GT configurations) and thus this model is preferred for the analysis here. Note that model B is preferred to model A in this particular context as, based on past observations from the authors of Rolls-Royce combustors, it provides a better balance of the reactive term $\mathcal{R}$ in Eq.~\eqref{eq:cvar_1}, which remains of leading order for $\Delta \gg \delta_{th}$. 
  
  When model B of Table~\ref{tab:SDR} is used without flame speed correction, the flame is observed to assume the unwanted V-shape configuration. This gives scope to investigate on whether the correction in Eq.~\eqref{eq:CorrectedOmega} can trigger a switch to the correct flame shape without resulting in a too-weak flame at the same time. Indeed, a too-strong limitation of the integrated reaction rate can result in an unphysically weak flame (as observed for Model A of Table~\ref{tab:SDR}) and also suggests limitations of the proposed correction.
  This analysis is, here, carried out in a first attempt, in a qualitative manner due to limited computational resources and simulation time to assess the capabilities of the proposed speed corrector at GT conditions.
  Since the multi-grid analysis performed for the quasi-laminar case in Section~\ref{sec:Results} is not performed here, the correct level of variance cannot be rigorously verified. Nonetheless, the information presented in the next section still provides useful insights on the capabilities of the speed correction, also considering the wide use of model B in academia and industry.

  \subsubsection{LES results for the gas turbine configuration} \label{sec:Results_3}
 For a similar injector and configuration, \cite{LangellaHBVSZ2020} showed that the transition from M-shape to V-shape occurs when the flame, due to local periodic oscillations, moves to regions upstream of richer mixture and lower velocities created by the temporary formation of an inner recirculation zone (IRZ). Although this was identified as a secondary mechanism (the primary mechanism being related to vorticity from the inlet), it was shown that it is sufficient to trigger the transition when the SGS variance is very low. 
 It is thus interesting to understand if the generalised correction model discussed in Section~\ref{sec:PPflame_description} can revert the flame back to the M-shape configuration at least from a qualitative point of view. This is analysed here.
 It is worth recalling that the function $f(\Delta)$ does not affect directly the SGS variance as the reaction term $\mathcal{R}$ in Eq.~\eqref{eq:cvar_1} is not scaled by $f$. The correction function can only affect the SGS variance indirectly, as a different flame speed implies a different reaction rate, thus different temperature, density and velocity.
 
\begin{figure}[ptb]
   	\centering
    \includegraphics[width=\columnwidth]{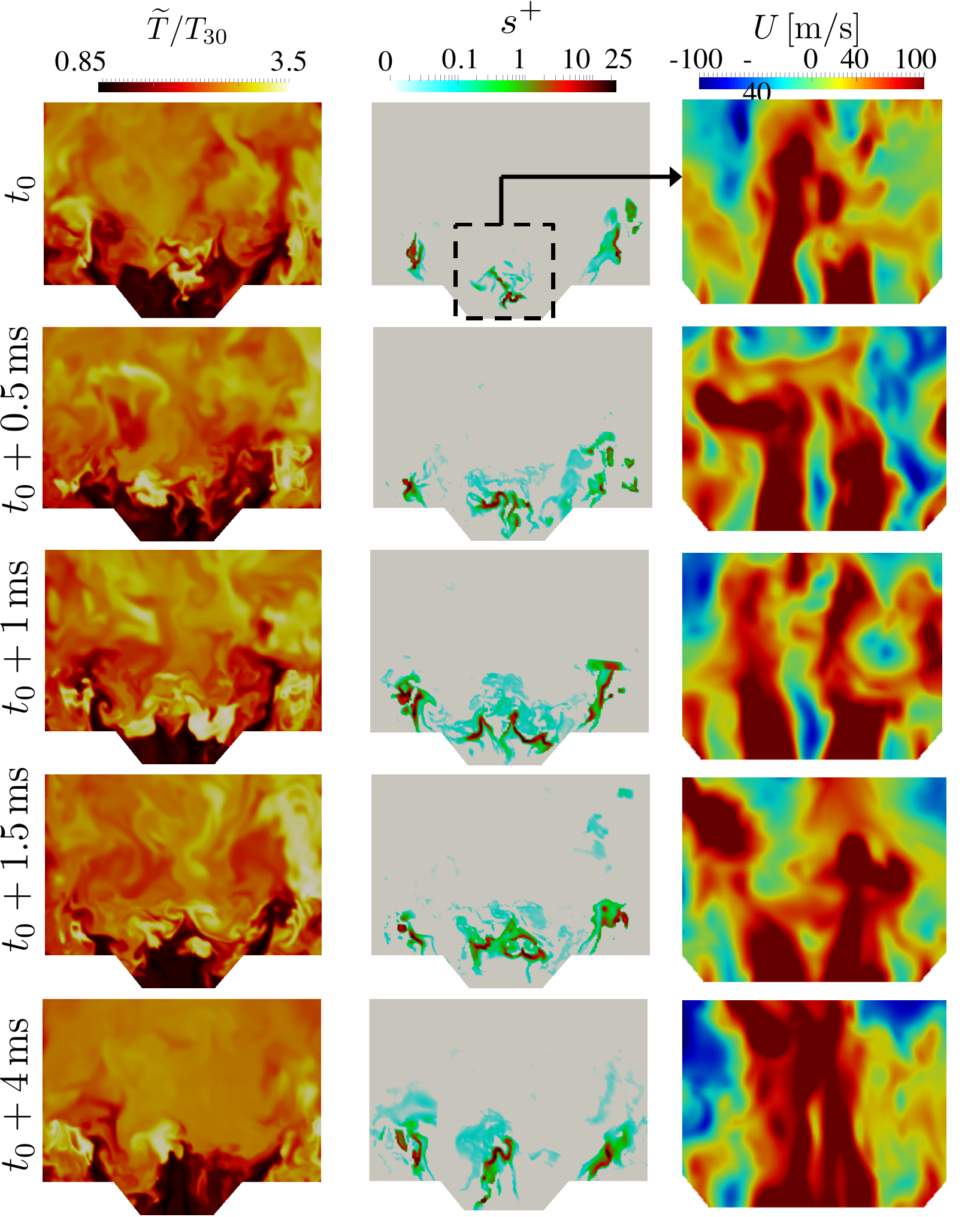}
	\caption{Mid-plane contours of temperature, normalised flame speed and axial velocity (truncated and zoomed in the indicated region) showing transition from a V-flame to M-flame after the flame speed corrector is activated.}	\label{fig:Alecsys_timeSeq}
\end{figure}

When no flame speed correction is used, the flame stabilises in the V-shape configuration discussed earlier, as shown by midplane contours of temperature, normalised flame speed and axial velocity of Fig.~\ref{fig:Alecsys_timeSeq} for a time $t=t_0$ taken after 
6 ms of statistically steady state (about 8 flow-through times)
Unlike the premixed flame configuration studied in Section~\ref{sec:Results}, the flame does not exhibit flashback here using model B, which is partly due to the high inlet speeds and partly because the turbulent production term $\mathcal{P}$ at GT regimes is more comparable to the reaction term $\mathcal{R}$ in Eq.~\eqref{eq:cvar_1} than for the flamelet regimes, thus the assumptions underlying model B are more robust, for this case. Nevertheless, the reaction source $\mathcal{R}$ remains of leading order for $\Delta/\delta_{th} >1$~\citep{LangellaS2016}, which is why the assumptions for model A of Table~\ref{tab:SDR} remain unsatisfactory. As mentioned, an unrealistic very weak flame was observed in this case (not shown). 
 
At time $t>t_0$ the flame speed correction model is activated. As can be observed from Fig.~\ref{fig:Alecsys_timeSeq}, the pilot flame at subsequent times starts to move to downstream positions (more penetrating jet), while the main flame remains at about the same position. It is worth pointing out that the normalised flame speed $s^+$ first decreases at time $t_0+0.5\,{\rm ms}$ to then increase again as the flame moves downstream. This is because the flame is moving to a region of higher speed surrounding the IRZ, indicated by the negative axial velocities observed near the inlet in the velocity contour of Fig.~\ref{fig:Alecsys_timeSeq}.  The IRZ starts to disappear after time $t_0+1\,{\rm ms}$ and the flame then continues to move downstream at following times. From time $t_0+1.5\,\rm{ms}$, the flame enters a periodic oscillation mode with extreme positions observed for times $t_0+1.5\,{\rm ms}$ and $t_0+4\,{\rm ms}$, corresponding to frequency of 400 Hz, which is similar to what was observed in~\cite{LangellaHBVSZ2020}. The pilot jet opens and closes during this time as can also be observed from the axial velocity contour, and this also was observed in~\cite{LangellaHBVSZ2020}.
The physics underlying this oscillatory mechanism is not the objective of this paper, and interested readers can find additional details in~\cite{LangellaHBVSZ2020}. However, the results shown here demonstrate that, at least from a qualitative point of view, the introduction of the flame speed corrector is able to correct the reactive flow prediction and achieve the expected configuration. 
%

\begin{figure}[htb]
   	\centering
    \includegraphics[width=0.5\columnwidth]{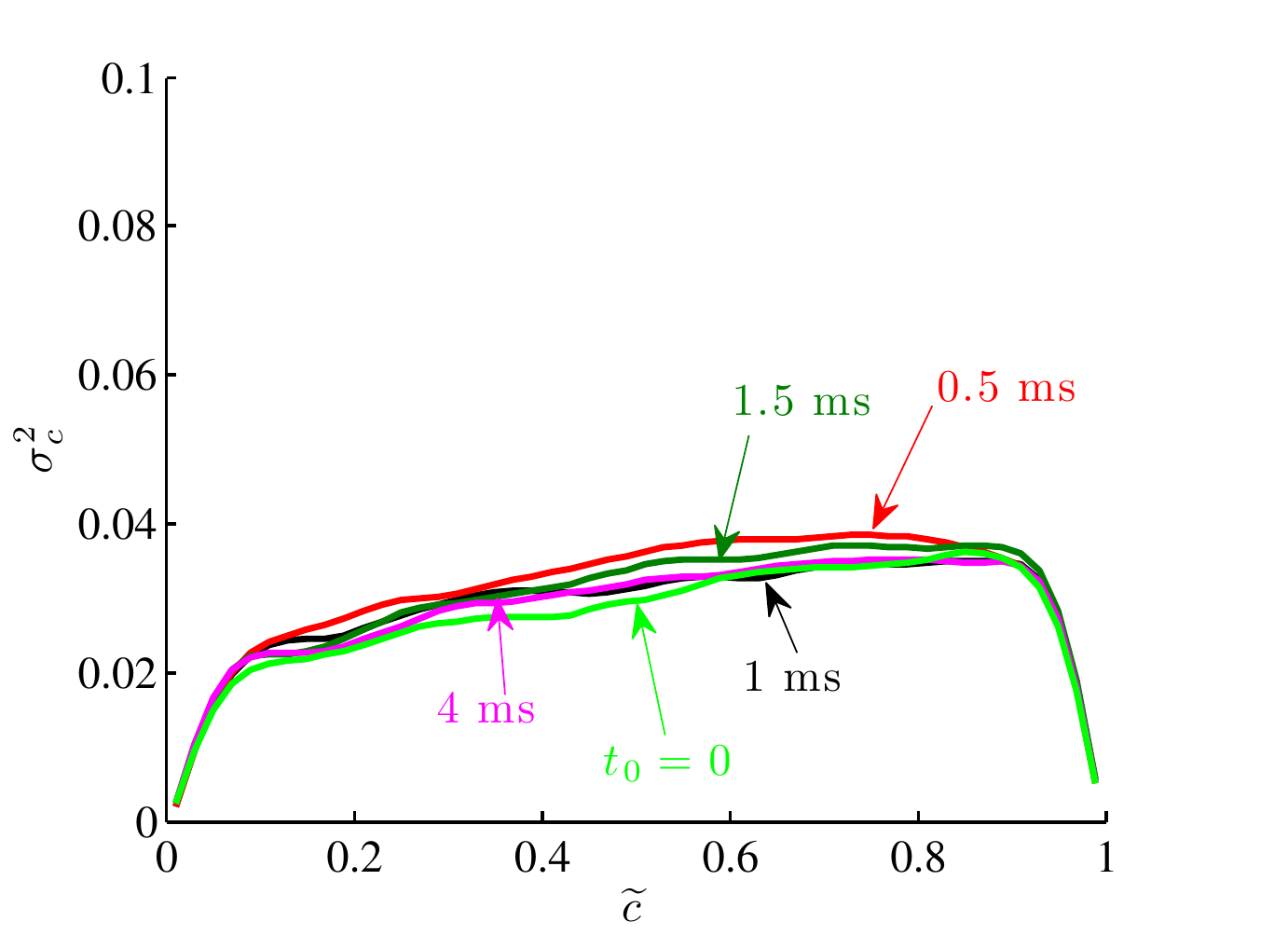}
	\caption{Midlplane contours of temperature, normalised flame speed and SGS variance of scaled progress variable showing transition from a V-flame to M-flame after the flame speed corrector is activated.}	\label{fig:Alecsys_condVar}
\end{figure}

%
One final remark concerns the role of the SGS variance. The conditional mean of $\sigma^2_c$ is shown in Fig.~\ref{fig:Alecsys_condVar} for the time sequence shown in Fig.~\ref{fig:Alecsys_timeSeq}. As observed for the premixed case, the variance does not significantly change after the introduction of the correction for the flame speed. During the time sequence, the SGS variance is first observed to increase at time $t=t_0+0.5\,{\rm ms}$ as a result of the increased reaction rate in the new position downstream (recall that the term $\mathcal{R}$ in Eq.~\eqref{eq:cvar_1} is not affected by the flame speed correction), after which it returns to values similar to those observed before the correction is applied. Since the same transition to the M-flame can be observed when the SGS variance increases, e.g. by switching to model C of Table~\ref{tab:SDR} as discussed in~\cite{LangellaHBVSZ2020}, 
some concerns on the meaning and role of the SGS variance arise. The fundamental question is whether the right value of $\sigma^2_c$ is the one observed in Fig.~\ref{fig:Alecsys_condVar}, or that one would obtain by varying the subgrid SDR model constants. In~\cite{LangellaHBVSZ2020} and \cite{SemlitschHLSD2019}, this second step was automatically achieved by using a dynamic evaluation of the subgrid SDR constants, and thus the variance could have been artificially increased by the dynamic process. On the contrary, the SGS variance obtained using the flame speed correction in this work is more physically related to the consumption speed, and thus may be a better candidate to represent the actual subgrid fluctuation level of $\widetilde{c}$. Additional investigations are necessary to better understand the role of SGS variance under dynamic evaluation of the combustion constant, which is the objective of future studies.



\section{Conclusions}\label{sec:Conclusions}
Large eddy simulations for premixed and partially premixed combustion have been presented in this work to assess the performance of a novel model to correct the flame consumption speed in presumed-PDF based models. The $\beta$-PDF is in particular known to overestimate the reaction rate under quasi-laminar conditions. This behaviour is discussed in this work and related to an incorrect prediction of the SGS variance by the $\beta$-PDF for cases where the filter width is larger than the laminar flame thickness. In this case, in fact, the SGS variance increases to signify the increased probability of heat release oscillations at the SGS level for a larger cell volume. However this increase is limited. The model tested in this work shows the ability to recover the correct flame speed behaviour without directly affecting the SGS variance and its intricate balance between reaction source, turbulence production and dissipation.

The first simulation concerns a premixed flame in the wrinkled flamelet regime of the premixed combustion diagram. 
Results are first obtained for a grid whose size does not require the use of the flame speed correction and compared to experimental data for reference. In the second step, the same numerical setup corresponding to the flame having a good match with experimental data is used on a coarser grid having filter size three times the size of the laminar flame thickness. In this case, it is shown that the flame consumption speed increases leading to a much shorter flame. The use of the correction model brings the flame back to the correct position, demonstrating the effectiveness of the model.

A generalised model is then proposed in an algebraic form to be used for a partially-premixed, high pressure gas turbine combustor.  When the correction model is employed, the expected oscillatory M-flame configuration is predicted in the GT configuration, while an incorrect V-shape is predicted otherwise. It is also shown that this model is weakly sensitive to fuel and operative conditions, thus could be generalized to different conditions, and it is easy to implement in any code due to its algebraic form.

Finally, it is observed that the proposed model does not affect the SGS variance, which leads to some unanswered questions about the role of this quantity in a LES. In fact, the flame consumption speed can be equivalently corrected by adding extra variance (e.g. by varying the constants in the SDR model or using dynamic approaches), but this may act as compensation of effects. Instead, the SGS variance obtained using the proposed modelling is more intrinsically related to the consumption speed which is a physical quantity, and thus may be a better candidate to represent the actual subgrid fluctuation levels. Future work will investigate more finely this role of the SGS variance and the interplay between reactive source, turbulent production and scalar dissipation rate
in the context of this flame consumption speed correction.

In conclusion, the results presented in this paper show that the burning speed correction can increase the accuracy of the flamelet model when this is coupled with a presumed-PDF approach, and also suggest that the overestimation of the burning rates commonly observed for certain combustion regimes may be a consequence of volume integration rather than of the modelling underlying assumptions. In principle, this correction can be applied to any flamelet-like model such as the Flamelet Generated Manifold to ensure that the consumption speed is correctly predicted when a presumed-PDF approach is applied. However, additional tests need to be conducted at different combustion regimes, including spray effects, before this correction can be generalised. This is left for future work.

\bibliographystyle{apalike_modif}

\begin{thebibliography}{}

\bibitem[Che, 2002]{Chem1d_code}
 2002.
\newblock {CHEM1d} 2002. {A} one dimensional flame code. {E}indhoven
  {U}niversity of {T}echnology.
\newblock http://www.combustion.tue.nl/chem1d.

\bibitem[COR, 2020]{CORNET_URL}
 2020.
\newblock Aircraft noise. getting to the core of the issue with cornet.
\newblock
  https://www.cleansky.eu/aircraft-noise-getting-to-the-core-of-the-issue-with-cornet.

\bibitem[Abdel-Gayed and Bradley, 1985]{AbdelB1985}
Abdel-Gayed, RG and Bradley, D 1985.
\newblock Criteria for turbulent propagation limits of premixed flames.
\newblock {\em Combust. Flame}, 62:61--68.

\bibitem[Abdel-Gayed et~al., 1987]{AbdelBL1987}
Abdel-Gayed, RG, Bradley, D, and Lawes, M 1987.
\newblock Turbulent burning velocities: a general correlation in terms of
  straining rates.
\newblock {\em Proc. Royal Soc. Lond. A}, 414:389--413.

\bibitem[Allauddin et~al., 2017]{AllauddinKPC2017}
Allauddin, U, Klein, M, Pfitzner, M, and Chakraborty, N 2017.
\newblock A priori and a posteriori analyses of algebraic flame surface density
  modeling in the context of large eddy simulation of turbulent premixed
  combustion.
\newblock {\em Numer. Heat Transfer, Part A: Appl.}, 71(2):153--171.

\bibitem[Anand et~al., 2013]{AnandESZZ2013}
Anand, MS, Eggels, R, Staufer, M, Zedda, M, and Zhu, J 2013.
\newblock An advanced unstructured-grid finitevolume design system for gas
  turbine combustion analysis.
\newblock {\em Proc. {ASME} Gas Turbine India (GTINDIA2013-3537)}.

\bibitem[Benard et~al., 2019]{BenardLMM2019}
Benard, P, Lartigue, G, Moureau, V, and Mercier, R 2019.
\newblock Large-{E}ddy simulation of the lean-premixed {PRECCINSTA} burner with
  wall heat loss.
\newblock {\em Proc. Combust. Inst.}, 37:5233--5243.

\bibitem[Bisetti et~al., 2009]{BisettiCCH2009}
Bisetti, F, Chen, JY, Chen, JH, and Hawkes, ER 2009.
\newblock Differential diffusion effects during the ignition of a thermally
  stratified premixed hydrogen–air mixture subject to turbulence.
\newblock {\em Proc. Combust. Inst.}, 32:1465--1472.

\bibitem[Borghi, 1990]{Borghi1990}
Borghi, R 1990.
\newblock Turbulent premixed combustion: Further discussions on the scales of
  fluctuations.
\newblock {\em Combust. Flame}, 80(304-312).

\bibitem[Bray, 2016]{Bray2016}
Bray, K 2016.
\newblock Laminar flamelets in turbulent combustion modeling.
\newblock {\em Combust. Sci. Technol.}, 188(9):1372--1375.

\bibitem[Bray et~al., 2006]{BrayCLS2006}
Bray, KNC, Champion, M, Libby, PA, and Swaminathan, N 2006.
\newblock Finite rate chemistry and presumed {PDF} models for premixed
  turbulent combustion.
\newblock {\em Combust. Flame}, 146:665--673.

\bibitem[Bray et~al., 1985]{BrayLM1985}
Bray, KNC, Libby, PA, and Moss, JB 1985.
\newblock Unified modeling approach for premixed turbulent combustion -- part
  {I}: General formulation.
\newblock {\em Combust. Flame}, 61:87--102.

\bibitem[Bressloff, 2011]{Bressloff2011}
Bressloff, NW 2011.
\newblock A parallel pressure implicit splitting of operators algorithm applied
  to flows at all speeds.
\newblock {\em Int. J. Numer. Methods Fluids}, 36:497--518.

\bibitem[Chai and Mahesh, 2012]{ChaiM2010}
Chai, X and Mahesh, K 2012.
\newblock Dynamic k-equation model for large eddy simulation of compressible
  flow.
\newblock {\em J. Fluid Mech.}, 699:385--413.

\bibitem[Chakraborty and Cant, 2009]{ChakrabortyC2009}
Chakraborty, N and Cant, RS 2009.
\newblock Effects of lewis number on scalar transport in turbulent premixed
  flames.
\newblock {\em Phys. Fluids}, 21:035110.

\bibitem[Charlette et~al., 2002]{Charlette2002}
Charlette, F, Meneveau, C, and Veynante, D 2002.
\newblock A power-law flame wrinkling model for {LES} of premixed turbulent
  combustion, {P}art {I}: Non-dynamic formulation and initial tests.
\newblock {\em Combustion and Flame}, 131(1--2):159--180.

\bibitem[Chen et~al., 2018]{ChenDRLS2018}
Chen, ZX, Doan, NAK, Ruan, S, Langella, I, and Swaminathan, N 2018.
\newblock A priori investigation of subgrid correlation of mixture fraction and
  progress variable in partially premixed flames.
\newblock {\em Combust. Theory Model.}, 22(5):862--882.

\bibitem[Chen et~al., 2020]{ChenLBS2019}
Chen, ZX, Langella, I, Barlow, RS, and Swaminathan, N 2020.
\newblock Prediction of local extinctions in piloted jet flames with
  inhomogeneous inlets using unstrained flamelets.
\newblock {\em Combust. Flame}, 212:415--432.

\bibitem[Clark et~al., 1979]{ClarkFR1979}
Clark, RA, Ferziger, JH, and Reynolds, WC 1979.
\newblock Evaluation of subgrid-scale models using an accurately simulated
  turbulent flow.
\newblock {\em J. Fluid Mech.}, 91:1–16.

\bibitem[Colin et~al., 2000]{ColinDVP2000}
Colin, O, Ducros, F, Veynante, D, and Poinsot, TJ 2000.
\newblock A thickened flame model for large eddy simulations of turbulent
  premixed combustion.
\newblock {\em Phys. Fluids}, 12:1843.

\bibitem[Correa, 1998]{Correa1998}
Correa, SM 1998.
\newblock A review of ${\rm no_x}$ formation under gas-turbine combustion
  conditions.
\newblock {\em Proc. Combust. Inst.}, 28:1793--1807.

\bibitem[Dagaut and Cathonnet, 2006]{DagautC2006}
Dagaut, P and Cathonnet, M 2006.
\newblock The ignition, oxidation, and combustion of kerosene: A review of
  experimental and kinetic modeling.
\newblock {\em Prog. Energy Combust. Sci.}, 32:48--92.

\bibitem[Darbyshire and Swaminathan, 2012]{DarbyshireS2012}
Darbyshire, OR and Swaminathan, N 2012.
\newblock A presumed joint pdf model for turbulent combustion with varying
  equivalence ratio.
\newblock {\em Combust. Sci. Technol.}, 184:2036--2067.

\bibitem[Darbyshire et~al., 2010]{DarbyshireSH2010}
Darbyshire, OR, Swaminathan, N, and Hochgreb, S 2010.
\newblock The effects of small-scale mixing models on the prediction of
  turbulent premixed and stratified combustion.
\newblock {\em Combust. Sci. Technol.}, 182:1141--1170.

\bibitem[Davis, 1970]{Abramo70}
Davis, PJ 1970.
\newblock Gamma functions and related functions.
\newblock In Abramowitz, M and Stegun, IA, editors, {\em Handbook of
  mathematical functions}. Dover Publications Inc., New York.

\bibitem[Dinesh et~al., 2016]{DineshSLOT2016}
Dinesh, KKJR, Shalaby, H, Luo, KH, van Oijen, JA, and Th\'evenin, D 2016.
\newblock High hydrogen content syngas fuel burning in lean premixed spherical
  flames at elevated pressures: Effects of preferential diffusion.
\newblock {\em Int. J. Hydrogen Combust.}, 41:18231--18249.

\bibitem[Doan et~al., 2017]{DoanSC2017}
Doan, NAK, Swaminathan, N, and Chakraborty, N 2017.
\newblock Multiscale analysis of turbulence-flame interaction in premixed
  flames.
\newblock {\em Proc. Combust. Inst.}, 36(2):1929--1935.

\bibitem[Domingo et~al., 2005]{DomingoVPH2005}
Domingo, P, Vervisch, L, Payet, S, and Hauguel, R 2005.
\newblock {DNS} of a premixed turbulent {V} flame and {LES} of a ducted flame
  using a {FSD-PDF} subgrid scale closure with {FPI}-tabulated chemistry.
\newblock {\em Combust. Flame}, 143:566--586.

\bibitem[Donini et~al., 2015]{DoniniBOG2015}
Donini, A, Bastiaans, RJM, van Oijen, JA, and de~Goey, LPH 2015.
\newblock Differential diffusion effects inclusion with flamelet generated
  manifold for the modeling of stratified premixed cooled flames.
\newblock {\em Proc. Combust. Inst.}, 35:831--837.

\bibitem[Donini et~al., 2017]{DoniniBOG2017}
Donini, A, Bastiaans, RJM, van Oijen, JA, and de~Goey, LPH 2017.
\newblock A 5-{D} implementation of {FGM} for the large eddy simulation of a
  stratified swirled flame with heat loss in a gas turbine combustor.
\newblock {\em Flow Turbul. Combust.}, 98:887--922.

\bibitem[Doormaal and Raithby, 1984]{VanDoormaalR1984}
Doormaal, JPV and Raithby, GD 1984.
\newblock Enhancements of the simple method for predicting incompressible fluid
  flows.
\newblock {\em Numer. Heat Transfer}, 7:147--163.

\bibitem[Dunn et~al., 2010]{DunnMBB2010}
Dunn, MJ, Masri, AR, Bilger, RW, and Barlow, RS 2010.
\newblock Finite rate chemistry effects in highly sheared turbulent premixed
  flames.
\newblock {\em Flow Turbulence Combust.}, 85:621--648.

\bibitem[Dunstan et~al., 2013]{DunstanMCS2013}
Dunstan, T, Minamoto, Y, Chakraborty, N, and Swaminathan, N 2013.
\newblock Scalar dissipation rate modelling for large eddy simulation of
  turbulent premixed flames.
\newblock {\em Proc. Combust. Inst.}, 34:1193--1201.

\bibitem[Fiorina et~al., 2003]{FiorinaBGTCD2003}
Fiorina, B, Baron, R, Gicquel, O, Thevenin, D, Carpentier, S, and Darabiha, N
  2003.
\newblock Modelling non-adiabatic partially premixed flames using
  flame-prolongation of {ILDM}.
\newblock {\em Combust. Theory Model.}, 7:449--470.

\bibitem[Fiorina et~al., 2015]{FiorinaVC2015}
Fiorina, B, Veynante, D, and Candel, S 2015.
\newblock Modeling combustion chemistry in large eddy simulation of turbulent
  flames.
\newblock {\em Flow Turbul. Combust.}, 94:3--42.

\bibitem[Fiorina et~al., 2010]{FiorinaVADGV2010}
Fiorina, B, Vicquelin, R, Auzillon, P, Darabiha, N, Gicquel, O, and Veynante, D
  2010.
\newblock A filtered tabulated chemistry model for les of premixed combustion.
\newblock {\em Combust. Flame}, 157:465--475.

\bibitem[Furukawa et~al., 2016]{FurukawaYW2016}
Furukawa, J, Yoshida, Y, and Williams, FA 2016.
\newblock Structures of methane-air and propane-air turbulent premixed {B}unsen
  flames.
\newblock {\em Combust. Sci. Technol.}, 188(9):1538--1564.

\bibitem[Galeazzo et~al., 2019]{GaleazzoSWHCF2019}
Galeazzo, FCC, Savard, B, Wang, H, Hawkes, ER, Chen, J, and Filho, G 2019.
\newblock Performance assessment of flamelet models in flame-resolved les of a
  high karlovitz methane/air stratified premixed jet flame.
\newblock {\em Proc. Combust. Inst.}, 37:2545--2553.

\bibitem[Galpin et~al., 2008]{GalpinNVACD2008}
Galpin, J, Naudin, A, Vervisch, L, Angelberger, C, Colin, O, and Domingo, P
  2008.
\newblock Large-eddy simulation of a fuel-lean premixed turbulent swirl-burner.
\newblock {\em Combust. Flame}, 155:247--266.

\bibitem[Gicquel et~al., 2012]{GicquelSP2012}
Gicquel, LYM, Staffelbach, G, and Poinsot, T 2012.
\newblock Large eddy simulations of gaseous flames in gas turbine combustion
  chambers.
\newblock {\em Prog. Energy Combust. Sci.}, 38:782--817.

\bibitem[Jarrin et~al., 2006]{JarrinBLP2006}
Jarrin, N, Benhamadouche, S, Laurence, D, and Prosser, R 2006.
\newblock A synthetic-eddy method for generating inflow conditions for large
  eddy simulations.
\newblock {\em J. Heat Fluid Flow}, 27:585--593.

\bibitem[Jin et~al., 2008]{JinGB2008}
Jin, B, Grout, R, and Bushe, WK 2008.
\newblock Conditional source-term estimation as a method for chemical closure
  in premixed turbulent reacting flow.
\newblock {\em Flow Turbul. Combust.}, 81(4):563--582.

\bibitem[Kee et~al., 1985]{KeeGSM1985}
Kee, RJ, Grcar, JF, Smooke, MD, and Miller, JA 1985.
\newblock {\em A fortran program for modeling steady laminar one-dimensional
  premixed flames}.
\newblock Report {N}o. SAND85-8240, Sandia National Labortories, CA, USA.

\bibitem[Klapdor et~al., 2013]{KlapdorMKJ2013}
Klapdor, EV, di~Mare, F, Kollmann, W, and Janicka, J 2013.
\newblock A compressible pressure-based solution algorithm for gas turbine
  combustion chambers using the pdf/fgm model.
\newblock {\em Flow Turbul. Combust.}, 91:209--247.

\bibitem[Klein et~al., 2016]{KleinCP2016}
Klein, M, Chakraborty, N, and Pfitzner, M 2016.
\newblock Analysis of the combined modelling of sub-grid transport and filtered
  flame propagation for premixed turbulent combustion.
\newblock {\em Flow Turbul. Combust.}, 96:921--938.

\bibitem[Kolla and Swaminathan, 2010]{KollaS2010a}
Kolla, H and Swaminathan, N 2010.
\newblock Strained flamelets for turbulent premixed flames {II}: Laboratory
  flame results.
\newblock {\em Combust. Flame}, 157:1274--1289.

\bibitem[Langella et~al., 2017]{LangellaCSS2018}
Langella, I, Chen, ZX, Swaminathan, N, and Sadasivuni, SK 2017.
\newblock Large-eddy simulation of reacting flows in industrial gas turbine
  combustor.
\newblock {\em J. Propul. Power}, 34:1269--1284.

\bibitem[Langella et~al., 2018]{LangellaDSP2018}
Langella, I, Doan, NAK, Swaminathan, N, and Pope, SB 2018.
\newblock Study of subgrid-scale velocity models for reacting and nonreacting
  flows.
\newblock {\em Phys. Rev. Fluids}, 3.

\bibitem[Langella et~al., 2020]{LangellaHBVSZ2020}
Langella, I, Heinze, J, Behrendt, T, Voigt, L, Swaminathan, N, and Zedda, M
  2020.
\newblock Turbulent flame shape switching at conditions relevant for gas
  turbines.
\newblock {\em J. Eng. Gas Turbines Power}, 142. {P}aper {N}o GTP-19-1385(1).

\bibitem[Langella and Swaminathan, 2016]{LangellaS2016}
Langella, I and Swaminathan, N 2016.
\newblock Unstrained and strained flamelets for {LES} of premixed combustion.
\newblock {\em Combust. Theory Model.}, 20:410--440.

\bibitem[Langella et~al., 2015]{LangellaSGC2015}
Langella, I, Swaminathan, N, Gao, Y, and Chakraborty, N 2015.
\newblock Assessment of dynamic closure for premixed combustion {LES}.
\newblock {\em Combust. Theory Model.}, 19:628--656.

\bibitem[Langella et~al., 2016a]{LangellaSP2016}
Langella, I, Swaminathan, N, and Pitz, RW 2016a.
\newblock Application of unstrained flamelet {SGS} closure for multi-regime
  premixed combustion.
\newblock {\em Combust. Flame}, 173:161--178.

\bibitem[Langella et~al., 2016b]{LangellaSWF2016}
Langella, I, Swaminathan, N, Williams, FA, and Furukawa, J 2016b.
\newblock Large-eddy simulation of premixed combustion in the
  corrugated-flamelet regime.
\newblock {\em Combust. Sci. Technol.}, 188(9):1565--1591.

\bibitem[Lecocq et~al., 2010]{LecocqRCV2010}
Lecocq, G, Richard, S, Colin, O, and Vervisch, L 2010.
\newblock Gradient and counter-gradient modeling in premixed flames:
  theoretical study and application to the les of a lean premixed turbulent
  swirl-burner.
\newblock {\em Combust. Sci. and Tech.}, 182:465--479.

\bibitem[Lecocq et~al., 2011]{LecocqRCV2011}
Lecocq, G, Richard, S, Colin, O, and Vervisch, L 2011.
\newblock Hybrid presumed pdf and flame surface density approaches for
  large-eddy simulation of premixed turbulent combustion: Part 1: Formalism and
  simulation of a quasi-steady burner.
\newblock {\em Combust. Flame}, 158:1201--1214.

\bibitem[Libby and Bray, 1981]{LibbyB1981}
Libby, PA and Bray, KNC 1981.
\newblock Countergradient diffusion in premixed turbulent flames.
\newblock {\em {AIAA} J.}, 19:205--213.

\bibitem[Libby and Williams, 2000]{LibbyW2000}
Libby, PA and Williams, FA 2000.
\newblock Presumed pdf analysis of partially premixed turbulent combustion.
\newblock {\em Combust. Sci. Technol.}, 161:359--390.

\bibitem[Lieuwen and Yang, 2005]{LieuwenY2005}
Lieuwen, T and Yang, V 2005.
\newblock Combustion instabilities in gs turbine engines: operational
  experience, fundamental mechanisms and modeling.
\newblock In {\em Progress in Astronautics and Aeronautics}, volume 210. AIAA,
  American Institute of Aeronautics and Astronautics.

\bibitem[Lilly, 1967]{Lilly1967}
Lilly, DK 1967.
\newblock The representation of small-scale turbulence in numerical simulation
  experiments.
\newblock In Goldstine, HH, editor, {\em Proceedings of the IBM Scientific
  Computing Symposium on Environmental Sciences}, number 320-1951, pages
  195--210. IBM.

\bibitem[Lipatnikov and Sabelnikov, 2020]{LipatnikovS2021b}
Lipatnikov, AN and Sabelnikov, VA 2020.
\newblock An extended flamelet-based presumed probability density function for
  predicting mean concentrations of various species in premixed turbulent
  flames.
\newblock {\em Int. J. Hydrogen Energy}, 45:31162--31178.

\bibitem[Lipatnikov et~al., 2020]{LipatnikovSHSI2020}
Lipatnikov, AN, Sabelnikov, VA, Hernández-Pérez, FE, Song, W, and Im, HG
  2020.
\newblock A priori dns study of applicability of flamelet concept to predicting
  mean concentrations of species in turbulent premixed flames at various
  karlovitz numbers.
\newblock {\em Combust. Flame}, 222:370--382.

\bibitem[Lipatnikov et~al., 2021]{LipatnikovSHSI2021}
Lipatnikov, AN, Sabelnikov, VA, Hernández-Pérez, FE, Song, W, and Im, HG
  2021.
\newblock Prediction of mean radical concentrations in lean hydrogen-air
  turbulent flames at different karlovitz numbers adopting a newly extended
  flamelet-based presumed {PDF}.
\newblock {\em Combust. Flame}, 226:248--259.

\bibitem[Lysenko and Ertesv{\aa}g, 2018]{LysenkoE2018}
Lysenko, DA and Ertesv{\aa}g, IS 2018.
\newblock Reynolds-averaged, scale-adaptive and large-eddy simulations of
  premixed bluff-body combustion using the eddy dissipation concept.
\newblock {\em Flow Turbul. Combust.}, 100:721--768.

\bibitem[Ma and Roekaerts, 2016]{MaR2016}
Ma, L and Roekaerts, DJE 2016.
\newblock Modeling of spray jet flame under mild condition with non-adiabatic
  fgm and a new conditional droplet injection model.
\newblock {\em Combust. Flame}, 165:402--423.

\bibitem[Mercier et~al., 2014]{MercierAMDGVF2014}
Mercier, R, Auzillon, P, Moureau, V, Darabiha, N, Gicquel, O, Veynante, D, and
  Fiorina, B 2014.
\newblock {LES} modeling of the impact of heat losses and differential
  diffusion on turbulent stratified flame propagation: application to the {TU}
  {D}armstadt stratified flame.
\newblock {\em Flow Turbul. Combust.}, 93:349--381.

\bibitem[Moureau et~al., 2011]{MoureauDV2011}
Moureau, V, Domingo, P, and Vervisch, L 2011.
\newblock From large-eddy simulation to direct numerical simulation of a lean
  premixed swirl flame: Filtered laminar flame-{PDF}.
\newblock {\em Combust. Flame}, 158:1340--1357.

\bibitem[Nambully et~al., 2014]{NambullyDMV2014}
Nambully, S, Domingo, P, Moureau, V, and Vervisch, L 2014.
\newblock A filtered-laminar-flame {PDF} sub-grid-scale closure for {LES} of
  premixed turbulent flames: {II}. application to a stratified bluff-body
  burner.
\newblock {\em Combustion and Flame}, 161(7):1775 -- 1791.

\bibitem[Nilsson et~al., 2019a]{NilssonLDSYB2019}
Nilsson, T, Langella, I, Doan, NAK, Swaminathan, N, Yu, R, and Bai, XS 2019a.
\newblock A priori analysis of sub-grid variance of a reactive scalar using dns
  data of high ka flames.
\newblock {\em Combust. Theory Model.}, 23:885--906.

\bibitem[Nilsson et~al., 2019b]{NillsonYDLSB2019}
Nilsson, T, Yu, R, Doan, NAK, Langella, I, Swaminathan, N, and Bai, XS 2019b.
\newblock Filtered reaction rate modelling in moderate and high {K}arlovitz
  number flames: an \textit{a Priori} analysis.
\newblock {\em Flow Turbul. Combust.}, 103(3):643--665.

\bibitem[Oijen et~al., 2016]{VanOijenDBTBdG2016}
Oijen, JAV, Donini, A, Bastiaans, RJM, ten Thije~Boonkkamp, JHM, and de~Goey,
  LPH 2016.
\newblock State-of-the-art in premixed combustion modeling using flamelet
  generated manifolds.
\newblock {\em Prog. Energy Combust. Sci.}, 57:30--74.

\bibitem[Ottino et~al., 2016]{OttinoFFBG2016}
Ottino, GM, Fancello, A, Falcone, M, Bastiaans, RJM, and de~Goey, LPH 2016.
\newblock Combustion modeling including heat loss using flamelet generated
  manifolds: a validation study in openfoam flow turbul. combust.
\newblock {\em Flow Turbul. Combust.}, 96:773--800.

\bibitem[Pierce and Moin, 2004]{PierceM2004}
Pierce, CD and Moin, P 2004.
\newblock Progress-variable approach for large-eddy simulation of non-premixed
  turbulent combustion.
\newblock {\em J. Fluid Mech.}, 504:73--97.

\bibitem[Piomelli and Balaras, 2002]{PiomelliB2002}
Piomelli, U and Balaras, E 2002.
\newblock Wall-layer models for large-eddy simulations.
\newblock {\em Annu. Rev. Fluid Mech.}, 34(1):349--374.

\bibitem[Pitsch, 2006]{Pitsch2006}
Pitsch, H 2006.
\newblock Large-eddy simulation of turbulent combustion.
\newblock {\em Annu. Rev. Fluid Mech.}, 38:453--482.

\bibitem[Poinsot et~al., 1991]{PoinsotVC1991}
Poinsot, T, Veynante, D, and Candel, S 1991.
\newblock Quenching processes and premixed turbulent combustion diagrams.
\newblock {\em J. Fluid Mech.}, 228:561--606.

\bibitem[Poinsot and Veynante, 2005]{PoinsotV_book}
Poinsot, TJ and Veynante, D 2005.
\newblock {\em Theoretical and numerical combustion}.
\newblock Edwards.

\bibitem[Pope, 2000]{Pope_book}
Pope, SB 2000.
\newblock {\em Turbulent Flows}.
\newblock Cambridge University Press.

\bibitem[Proch and Kempf, 2015]{ProchK2015}
Proch, F and Kempf, AM 2015.
\newblock Modeling heat loss effects in the large eddy simulation of a model
  gas turbine combustor with premixed flamelet generated manifolds.
\newblock {\em Proc. Combust. Inst.}, 35:3337--3345.

\bibitem[Ribert et~al., 2004]{RibertCP2004}
Ribert, G, Champion, M, and Plion, P 2004.
\newblock Modeling a turbulent reactive flow with variable equivalence ratio:
  application to the calculation of a reactive shear layer.
\newblock {\em Combust. Sci. Technol.}, 176:907--923.

\bibitem[Roberts et~al., 1993]{RobertsDDG1993}
Roberts, W, Driscoll, J, Drake, M, and Goss, L 1993.
\newblock Images of the quenching of a flame by a vortex---to quantify regimes
  of turbulent combustion.
\newblock {\em Combust. Flame}, 94:58--69.

\bibitem[Robin et~al., 2008]{RobinMCDRB2008}
Robin, V, Mura, A, Champion, M, Degardin, O, Renou, B, and Boukhalfa, M 2008.
\newblock Experimental and numerical analysis of stratified turbulent v-shaped
  flames.
\newblock {\em Combust. Flame}, 153:288--315.

\bibitem[Salehi and Bushe, 2010]{SalehiB2010}
Salehi, MM and Bushe, W 2010.
\newblock Presumed {PDF} modelling for {RANS} simulation of turbulent premixed
  flames.
\newblock {\em Combust. Theory Model.}, 14(3):381--403.

\bibitem[Salehi et~al., 2013]{SalehiBSG2013}
Salehi, MM, Bushe, WK, Shahbazian, N, and Groth, CPT 2013.
\newblock Modified laminar flamelet presumed probability density function for
  {LES} of premixed turbulent combustion.
\newblock {\em Proc. Combust. Inst}, 34(1203-1211).

\bibitem[Semlitsch et~al., 2019]{SemlitschHLSD2019}
Semlitsch, B, Hynes, T, Langella, I, Swaminathan, N, and Dowling, AP 2019.
\newblock Entropy and vorticity wave generation in realistic gas turbine
  combustors.
\newblock {\em J. Propul. Power}, 35(4):839--849.

\bibitem[Sitte et~al., 2021]{SitteAGMC2021}
Sitte, MP, d'Auzay, CT, Giusti, A, Mastorakos, E, and Chakraborty, N 2021.
\newblock A-priori validation of scalar dissipation rate models for turbulent
  non-premixed flames.
\newblock {\em Flow Turbul. Combust.}, 107:201--218.

\bibitem[Soli et~al., 2021]{SoliLC2021}
Soli, A, Langella, I, and Chen, ZX 2021.
\newblock Analysis of flame front breaks appearing in les of inhomogeneous jet
  flames using flamelets.
\newblock {\em Flow Turbul. Combust.}

\bibitem[Temme et~al., 2015]{TemmeWSD2015}
Temme, JE, Wabel, TM, Skiba, AW, and Driscoll, JF 2015.
\newblock {\em Measurements of premixed turbulent combustion regimes of high
  {R}eynolds number flames}.
\newblock 53$^{rd}$ {A}erospace {S}ciences {M}eeting, {AIAA} {P}aper 2015-0168.

\bibitem[Trisjono et~al., 2014]{TrisjonoKKP2014}
Trisjono, P, Kleinheinz, K, Kang, S, and Pitsch, H 2014.
\newblock Large eddy simulation of stratified and sheared flames of a premixed
  turbulent stratified flame burner using a flamelet model with heat loss.
\newblock {\em Flow Turbul. Combust.}, 92:201--235.

\bibitem[van Oijen and de~Goey, 2000]{VanOijenG2000}
van Oijen, J and de~Goey, L 2000.
\newblock Modelling of premixed laminar flames using flamelet-generated
  manifold.
\newblock {\em Combust. Sci. and Tech.}, 161:113--137.

\bibitem[Veynante et~al., 1997]{VeynanteTBM1997}
Veynante, D, Trouv\'{e}, A, Bray, KNC, and Mantel, T 1997.
\newblock Gradient and counter-gradient transport in turbulent premixed flames.
\newblock {\em J. Fluid Mech.}, 332:263--293.

\bibitem[Volpiani et~al., 2016]{VolpianiSV2016}
Volpiani, PS, Schmitt, T, and Veynante, D 2016.
\newblock \textit{A posteriori} tests of a dynamic thickened flame model for
  large eddy simulations of turbulent premixed combustion.
\newblock {\em Combustion and Flame}, 174:166 -- 178.

\bibitem[Vreman et~al., 2009]{VremanOG2009}
Vreman, AW, van Oijen, JA, and de~Goey, LPH 2009.
\newblock Subgrid scale modelling in large-eddy simulation of turbulent
  combustion using premixed flamelet chemistry.
\newblock {\em Flow Turb. Combust.}, 82:511--535.

\bibitem[Weller et~al., 1998]{Weller1998}
Weller, HG, Tabor, G, Jasak, H, and Fureby, C 1998.
\newblock A tensorial approach to computational continuum mechanics using
  object-oriented techniques.
\newblock {\em Computers Phys.}, 12(6):620--631.

\end{thebibliography}

%
\newpage
\clearpage
\begin{table}[bt]
\caption{Summary of models used for the SGS scalar dissipation rate of the progress variable. \RB{The model constant is $C_\alpha$ in for models A and B, and $\beta_c$ for model C. }} 
\label{tab:SDR}
\centering
\begin{tabular}{lccc}
\hline
Model  &  Reaction source in Eq.~\eqref{eq:cvar_1}  &  subgrid SDR & Model constant \\
\hline 
A       & \RB{Present}   & Eq.~\eqref{eq:linearRelax} & 0.5-1.0 \\
B       & \RB{Not present ($\mathcal{R}=0$)}    & Eq.~\eqref{eq:linearRelax} & 0.5-1.0\\
C       & \RB{Present}   & Eq.~\eqref{eq:SDRmodel} & 2.4-30\\
\hline
\end{tabular}
\end{table}
%



%
%

%

%

%

%

%

%

%

%

%
\clearpage
\end{document}